\documentclass[twocolumn]{aastex631}

\usepackage{makecell}
\usepackage{amsmath}
\usepackage{longtable}
\usepackage{xtab}
\usepackage{pifont}
\usepackage{hyperref}
\usepackage{footnote}
\usepackage{apjfonts, natbib} \newenvironment{tight_enumerate}{
\begin{enumerate}
  \setlength{\itemsep}{0pt}
  \setlength{\parskip}{0pt}
}{\end{enumerate}}

\begin{document}

%\title{An Automated Method for Detecting TDE Candidates with Machine Learning Assistance for the Wide Field Survey Telescope (WFST)}

\title{TTC: Transformer-based TDE Classifier for the Wide Field Survey Telescope (WFST)}

\correspondingauthor{Ranfang Zheng, Zheyu Lin, Xu Kong}
\email{rfzheng@mail.ustc.edu.cn}
% \correspondingauthor{Zheyu Lin}
\email{linzheyu@mail.ustc.edu.cn}
% \correspondingauthor{Xu Kong}
\email{xkong@ustc.edu.cn}

% \email{rfzheng@mail.ustc.edu.cn, linzheyu@mail.ustc.edu.cn, xkong@ustc.edu.cn}
% \correspondingauthor{Zheyu Lin}
% \email{linzheyu@mail.ustc.edu.cn}
% \correspondingauthor{Xu Kong}
% \email{xkong@ustc.edu.cn}

%email : rfzheng@mail.ustc.edu.cn, linzheyu@mail.ustc.edu.cn, xkong@ustc.edu.cn
\author[0009-0006-4622-1417]{Ranfang Zheng}
\affil{Department of Astronomy, University of Science and Technology, Hefei, 230026, China}
\affil{School of Astronomy and Space Sciences, University of Science and Technology of China, Hefei 230026, China}
% \email{rfzheng@mail.ustc.edu.cn}

\author[0000-0003-4959-1625]{Zheyu Lin}
\affil{Department of Astronomy, University of Science and Technology, Hefei, 230026, China}
\affil{School of Astronomy and Space Sciences, University of Science and Technology of China, Hefei 230026, China}
% \email{linzheyu@mail.ustc.edu.cn}

\author[0000-0002-7660-2273]{Xu Kong}
\affil{Department of Astronomy, University of Science and Technology, Hefei, 230026, China}
\affil{School of Astronomy and Space Sciences, University of Science and Technology of China, Hefei 230026, China}
\affil{Institute of Deep Space Sciences, Deep Space Exploration Laboratory, Hefei 230026, China}
% \email{xkong@ustc.edu.cn}

%\author{Jie Lin}
%\affil{Department of Astronomy, University of Science and Technology, Hefei, 230026, China}

\author{Dezheng Meng}
\affil{Department of Astronomy, University of Science and Technology, Hefei, 230026, China}
\affil{School of Astronomy and Space Sciences, University of Science and Technology of China, Hefei 230026, China}
%\email{dezhengmeng@mail.ustc.edu.cn}

\author{Zelin Xu}
\affil{Department of Astronomy, University of Science and Technology, Hefei, 230026, China}
\affil{School of Astronomy and Space Sciences, University of Science and Technology of China, Hefei 230026, China}
%\email{sa21022025@mail.ustc.edu.cn}

\author[0000-0003-4200-4432]{Lulu Fan}
\affil{Department of Astronomy, University of Science and Technology, Hefei, 230026, China}
\affil{School of Astronomy and Space Sciences, University of Science and Technology of China, Hefei 230026, China}
\affil{Institute of Deep Space Sciences, Deep Space Exploration Laboratory, Hefei 230026, China}

\author[0000-0002-9092-0593]{Ji-an Jiang}
\affil{Department of Astronomy, University of Science and Technology, Hefei, 230026, China}
\affil{School of Astronomy and Space Sciences, University of Science and Technology of China, Hefei 230026, China}
\affil{National Astronomical Observatory of Japan, National Institutes of Natural Sciences, Tokyo 181-8588, Japan}

\author[0000-0002-7152-3621]{Ning Jiang}
\affil{Department of Astronomy, University of Science and Technology, Hefei, 230026, China}
\affil{School of Astronomy and Space Sciences, University of Science and Technology of China, Hefei 230026, China}

\author[0000-0003-3965-6931]{Jie Lin}
\affil{Department of Astronomy, University of Science and Technology, Hefei, 230026, China}
\affil{School of Astronomy and Space Sciences, University of Science and Technology of China, Hefei 230026, China}

\author[0000-0002-1517-6792]{Tinggui Wang}
\affil{Department of Astronomy, University of Science and Technology, Hefei, 230026, China}
\affil{School of Astronomy and Space Sciences, University of Science and Technology of China, Hefei 230026, China}
\affil{Institute of Deep Space Sciences, Deep Space Exploration Laboratory, Hefei 230026, China}

\author[0000-0003-0694-8946]{Qingfeng Zhu}
\affil{Department of Astronomy, University of Science and Technology, Hefei, 230026, China}
\affil{School of Astronomy and Space Sciences, University of Science and Technology of China, Hefei 230026, China}
\affil{Institute of Deep Space Sciences, Deep Space Exploration Laboratory, Hefei 230026, China}

%============================================

\author{Feng Li}
\affil{State Key Laboratory of Particle Detection and Electronics, University of Science and Technology of China, Hefei 230026, China}
%\email{phonelee@ustc.edu.cn}

\author{Ming Liang}
\affil{National Optical Astronomy Observatory (NSF’s National Optical-Infrared Astronomy Research Laboratory) 950 N Cherry Ave. Tucson Arizona 85726, USA}
%\email{liangming@gmail.com}

\author{Hao Liu}
\affil{State Key Laboratory of Particle Detection and Electronics, University of Science and Technology of China, Hefei 230026, China}
%\email{lhnows@ustc.edu.cn}

\author{Zheng Lou}
\affil{Purple Mountain Observatory, Chinese Academy of Sciences, Nanjing 210023, China}
%\email{zhenglou@pmo.ac.cn}

\author[0000-0003-1297-6142]{Wentao Luo}
\affil{Institute of Deep Space Sciences, Deep Space Exploration Laboratory, Hefei 230026, China}
%\email{wtluo@ustc.edu.cn}

\author{Jinlong Tang}
\affil{Institute of Optics and Electronics, Chinese Academy of Sciences, Chengdu 610209, China}
%\email{ioetang@163.com}

\author{Hairen Wang}
\affil{Purple Mountain Observatory, Chinese Academy of Sciences, Nanjing 210023, China}
%\email{hairenwang@pmo.ac.cn}

\author[0000-0003-1617-2002]{Jian Wang}
\affil{Institute of Deep Space Sciences, Deep Space Exploration Laboratory, Hefei 230026, China}
\affil{State Key Laboratory of Particle Detection and Electronics, University of Science and Technology of China, Hefei 230026, China}
%\email{wangjian@ustc.edu.cn}

\author[0000-0002-1935-8104]{Yongquan Xue}
\affil{Department of Astronomy, University of Science and Technology, Hefei, 230026, China}
\affil{School of Astronomy and Space Sciences, University of Science and Technology of China, Hefei 230026, China}

\author{Dazhi Yao}
\affil{Purple Mountain Observatory, Chinese Academy of Sciences, Nanjing 210023, China}
%\email{yaodazhi@pmo.ac.cn}

\author[0000-0002-1463-9070]{Hong-fei Zhang}
\affil{State Key Laboratory of Particle Detection and Electronics, University of Science and Technology of China, Hefei 230026, China}
%\email{nghong@ustc.edu.cn}

\author[0000-0002-1330-2329]{Wen Zhao}
\affil{Department of Astronomy, University of Science and Technology, Hefei, 230026, China}
\affil{School of Astronomy and Space Sciences, University of Science and Technology of China, Hefei 230026, China}

\author[0000-0003-3728-9912]{Xianzhong Zheng}
\affil{Tsung-Dao Lee Institute and Key Laboratory for Particle Physics, Astrophysics and Cosmology, Ministry of Education, Shanghai Jiao Tong University, Shanghai, 201210, China}

\author{Yingxi Zuo}
\affil{Purple Mountain Observatory, Chinese Academy of Sciences, Nanjing 210023, China}
%\email{yxzuo@pmo.ac.cn}

%% automatically takes care of all commas and ``and"s between authors names.


%% Mark off the abstract in the ``abstract`` environment. 
\begin{abstract}

We propose the Transformer-based Tidal disruption events (TDE) Classifier (\texttt{TTC}), specifically designed to operate effectively with both real-time alert streams and archival data of the Wide Field Survey Telescope (WFST). 
% This approach provides a highly versatile, data-efficient, and accurate solution for the automated identification of tidal disruption flares based solely on their optical light curves. 
It aims to minimize the reliance on external catalogs and find TDE candidates from pure 
% maximizes the utilization of data 
% from the 
light curves, 
% thereby reducing reliance on external crossmatches, which provides us with a greater chance of 
which is more suitable for finding 
% identifying more 
TDEs in faint and distant galaxies. 
\texttt{TTC} consists of two key modules that can work independently: (1) A light curve parametric fitting module and (2) a Transformer (\texttt{Mgformer})-based classification network. The training of the latter module and evaluation for each module utilize a light curve dataset of 7413 spectroscopically classified transients from the Zwicky Transient Facility (ZTF). The \texttt{Mgformer}-based module is superior in performance and flexibility. Its representative recall and precision values are 0.79 and 0.76, respectively, and can be modified by adjusting the threshold. It can also efficiently find TDE candidates within 30 days from the first detection. For comparison, the parametric fitting module yields values of 0.72 and 0.40, respectively, while it is $>$10 times faster in average speed. Hence, the setup of modules allows a trade-off between performance and time, as well as precision and recall. \texttt{TTC} has successfully picked out all spectroscopically identified TDEs among ZTF transients in a real-time classification test, and selected $\sim$20 TDE candidates in the deep field survey data of WFST. The discovery rate will greatly increase once the differential database for the wide field survey is ready.

%When external host galaxy matching information is integrated into the alert stream, the proposed method achieves an accuracy exceeding 0.85 at most, with a recall rate consistently above 0.7. The approach has been rigorously tested on spectrally confirmed supernova data collected during the WFST pilot sky survey, yielding results that align well with expectations. 

%The method is now fully prepared for deployment in the official sky survey operations of WFST. We expect that this method can help to make more efficient use of light curve and small amount of external data for pre-selection, so as to save spectral observation consumption.

\end{abstract}



\section{Introduction} \label{sec:intro}

A tidal disruption event (TDE) occurs when a star passes sufficiently close to a black hole and gets torn apart by tidal forces. Approximately half of the stellar debris is bound to the black hole, forming an accretion disk, while the remaining material is ejected into space \citep{orgin_TDE,ori-tde-rev}. TDEs exhibit distinctive features across various electromagnetic wavelengths. While TDEs were initially discovered
% were  made 
% by the ROSAT All-Sky Survey 
in the soft X-ray sky survey in the late 1990s~\citep{1996A&A...309L..35B, 1999A&A...343..775K} through archived data, the optical band has taken the lead in the discovery number and rate in the recent decade thanks to the advanced wide-field optical sky surveys. 
%The X-ray band continued to dominate the discovery of TDEs in 2000s, although the discovery rate was only $\sim$1 per year (e.g., \citealt{xtde-2010,refId0}). 
Real-time detection in the optical band, combined with spectroscopic confirmation, has become the primary approach for identifying TDEs at present.
% In the recent decade, about 
Currently, more than one hundred optical-bright TDEs have been found,
% in the advanced wide-field optical sky surveys,
% such as Pan-STARRS (e.g., \citealt{Gezari_2012,ps1-11ah,ps16dtm}), ASAS-SN (e.g., \citealt{ASASSN-14ae,ASASSN-14li-dis}) and ZTF \citep[e.g.,][]{2019ApJ...872..198V,vanVelzen2021,Hammerstein_2023_ZTF_I,tde-catalog-2}, The current discovery rate is a few tens per year, 
mostly contributed by ZTF \citep[e.g.,][]{2019ApJ...872..198V,vanVelzen2021,Hammerstein_2023_ZTF_I,tde-catalog-2}. Their identification heavily relies on the unique spectroscopic features, including the rise and fall of a blue continuum, and the possibly adherent broad H, He and Bowen fluorescent emission lines \citep{zabludoff_2021}.

%光度函数，亮端暗端情况，adviced by Dr.Jiang
Although optical surveys currently play a dominant role in the discovery of TDEs, significant limitations remain due to observational depth, which hinder a comprehensive understanding of the TDE population. The faint-end cutoff of the TDE luminosity function, for instance, remains poorly constrained owing to the small number of detections, despite theoretical predictions suggesting a larger population of faint TDEs \citep{TDE-LF,ZTF-I-LF,tde-catalog-2}, and we have indeed found the existence of this group at present \citep{AT2020wey,AT2023clx2,AT2023clx}. In addition, current optically selected TDE samples are strongly biased toward quiescent host galaxies, with very few detections in star-forming environments. TDEs in such galaxies may be obscured by intense star formation activity or dust extinction \citep{atlas17jrp,md2,MIR-TDE}. Overcoming these limitations will require deeper observations and time-domain surveys with higher sensitivity and cadence to construct a more complete and unbiased sample.

% The growing number of optical TDE samples enables the analysis of their light curves on a statistically significant scale, offering insights into their common properties and behaviors. 
With the commissioning of next-generation sky survey facilities such as WFST \citep{wfst-sci} and Vera Rubin Observatory \citep{lsstsciencecollaboration2009lsstsciencebookversion,lsst-sci}, hundreds to thousands of optical-bright TDEs are predicted to be discovered each year \citep{LSST-TDE-Pro,wfst-tde}. This signals the 
% the study of optical TDEs is poised to enter the 
era of large-sample analysis, but also poses challenges on the spectroscopic resources. Automatic classification based on photometric data can serve as a plausible solution, as the current number of optical TDE samples is enough for training and testing.

Several machine learning (ML) algorithms have been specifically designed for multi-dimensional time-series classification. These models are either trained on simulated datasets like {\tt PLAsTiCC} or {\tt ELAsTiCC} (e.g., {\tt Avocado}:~\citealt{avocado}; {\tt ATAT}:~\citealt{ATAT}; {\tt ORACLE}:~\citealt{shah2025oraclerealtimehierarchicaldeeplearning}; \citealt{Tang_2025}), or trained on real survey datasets like 
% , such as that from 
ZTF (e.g., {\tt FLEET}:~\citealt{Gomez_2023_FLEET}; {\tt tdescore}:~\citealt{Stein_2024}; {\tt NEEDLE}:~\citealt{needle}; {\tt ALeRCE}:~\citealt{ALERCE_total,ALERCE-lc,ALERCE-IMG,ALERCE-Anomaly,ALERCE_tde}), while \citealt{TDE-early-stage} has focused on attempting to quickly identify TDEs based on ascending stage of light curves. Most of these algorithms incorporate additional external data, such as the properties of host galaxies. 
% , additional external data, including host galaxy colors, redshifts, AllWISE color and angular separations between transients and their hosts. 
In particular, redshift enables the determination of a source’s luminosity, and angular separation between the transient and host galaxy aids in classification as different types of transients have different preferences from the central to outskirt regions of the galaxy.

Although external data are indeed useful, they sometimes can exclude peculiar but important transients. Particularly, an off-center TDE, or a TDE occurring around a much smaller black hole than that in the center of galaxy, will be unfortunately excluded by a strict angular separation cut (e.g., $< 0.6^{\prime\prime}$ in \citealt{2019ApJ...872..198V}). Although they are
% could unfortunately exclude all these valuable events. is
% more accurately, TDEs around the 
% is 
estimated to occur at a rate of merely $\sim$1$\%$ of that of a central TDE, some reliable candidates have been identified \citep[e.g.,][]{off-cen-TDE4,off-cen-TDE5,off-cen-TDE-candidate}. They can signal the existence of intermediate-mass black holes (IMBHs) or runaway BHs that ejected from 
% alternatively, the off-center observed features may arise from 
% interactions involving binary supermassive black holes (SMBHs), including 
binary supermassive black hole (SMBHs) mergers or gravitational-wave recoil events
% , which can result in displaced or ejected SMBHs. 
(e.g., \citealt{off-cen-TDE3,off-cen-TDE1,off-cen-TDE2}),
% These off-center TDEs, are less affected by stellar contamination from dense nuclear regions. Consequently, they 
and provide a cleaner environment than dense nuclear regions for probing the BH accretion processes \citep{IMBH_rev,off-cen-TDE-candidate,2024tvd_2}. 
% Moreover, such events offer valuable insights into galaxy merger scenarios, the probe of nuclear star clusters (NSCs), and the growth processes of black hole mass \citep{}. 
Therefore, our goal is to 
% it is essential to 
develop a method for identifying TDE candidates solely based on their light curves, minimizing reliance on external cross-matching with other datasets.

% \textbf{
% % it
% % modelling works have predicted that a subset of TDEs may be significantly offset from the centers of their host galaxies , 
% Some reliable candidates have been identified \citep[e.g.,][]{off-cen-TDE4,off-cen-TDE5,off-cen-TDE-candidate}. }
% % Although the occurrence rate of such off-center TDEs is estimated to be only about 1$\%$ of nuclear TDEs, 
% Applying a strict angular separation cut (e.g., $< 0.6^{\prime\prime}$ in \citealt{2019ApJ...872..198V}) could unfortunately exclude all these valuable events.
% Therefore, it is essential to develop a method for identifying TDE candidates solely based on their light curves, minimizing reliance on external cross-matching with other datasets.

Transformer model excels at capturing long-range dependencies within time series data \citep{10150366,wen2023transformerstimeseriessurvey}. By leveraging positional encoding, transformer models can capture temporal information in a manner akin to Long Short-Term Memory (LSTM) and Gated Recurrent Units (GRUs),
% , \citealt{wen2023transformerstimeseriessurvey}), 
while also offering improved scalability, flexibility, and advantages in parallel computation.

While transformer models have shown promising results for transient source classification using simulated datasets, such as the {\tt ELAsTiCC} dataset \citep{ATAT}, their performance on real-world data remains largely untested. In this study, we introduce a variant of the transformer model, \texttt{Mgformer} \citep{MgFormer}, as the base algorithm of our TDE filter, we developed the Transformer-based TDE Classifier (\texttt{TTC}) to search for TDE candidates merely relying on the light curves, and apply this filter onto the ZTF and WFST
% observational 
light curve datasets.
% to identify and screen TDE candidates.

In addition to the machine learning approach, we also propose a complementary parametric method. This parametric framework enables a rapid reduction in the number of candidates requiring further inspection, thereby decreasing the computational cost and mitigating the screening workload of the machine learning pipeline. Moreover, the parametric method itself possesses intrinsic discriminatory power, allowing it to serve as an independent preliminary filter. The balance between the parametric and machine learning components can be flexibly adjusted according to the specific objectives and requirements of the task.

The structure of the paper is as follows. In \autoref{sec:data_collect} we briefly introduce the construction of our ZTF light curve dataset
% how \texttt{Lasair\footnote{\href{https://lasair-ztf.lsst.ac.uk/}{\url{https://lasair-ztf.lsst.ac.uk/}}}} API is used to obtain the ZTF light curve dataset 
for training and testing.
% , as well as the WFST pilot survey light curve dataset
In \autoref{sec:filter} we introduce our integrated search methodology in details. In \autoref{sec:supply-real-data}, we demonstrate the performance of this method on WFST survey data and present the representative TDE candidates that pass the filter. In \autoref{sec:discussion}, we discuss the ability of finding TDEs at early stages, the performance and computational efficiency of modules, and the special TDE types that will be taken into consideration in the future updates.
% , and the bonus ability of classifying supernovae (SNe).
% provide a more detailed analysis of the proposed method, and discuss specific types of TDE candidates of particular interest. 
In \autoref{sec:conclusion} we conclude the study with a summary of our findings.

\section{Construction of the ZTF light curve dataset} \label{sec:data_collect}

The light curves utilized in this paper are based on the PSF photometry on the differential images \citep{ZTF-data-process}. We introduce the construction of the ZTF light curve dataset in this section, and leave the introduction of the WFST light curve dataset in \autoref{sec:supply-real-data}.
% They consists of ZTF and WFST pilot survey : composed of two primary components: ZTF differential image light curves and WFST pilot survey data. 

To evaluate the accuracy and completeness of our TDE filter, our
% In addition to our primary target, TDE, the
dataset includes TDE and other 14 transient types, following the setting of {\tt PLAsTiCC} \citep{plasticc}. 
% namely SN Ia, SN Iax, SN Ia (91bg-like), SN II, SN Ib, SN Ic, SN Ibc, SN Ibn, SN IIP, SN IIn, SN IIb, SLSN\footnote{=superluminous supernova}-I, SLSN-II, and AGN.
% , comprising a total of 15 distinct classes
We further group these types into six categories, and use them for our later labelling:

\begin{tight_enumerate}
    \item TDE.
    \item SN Ia: SN Ia, SN Iax, and SN Ia (91bg-like).
    \item SN Ib/c: SN Ib, SN Ic, SN Ib/c, and SN Ibn.
    \item SN II: SN II, SN IIP, SN IIn, and SN IIb.
    \item SLSN (=superluminous supernova): SLSN-I and SLSN-II.
    \item AGN (=active galactic nucleus).
\end{tight_enumerate}

We limit the discovery date of the targets to the period from 2019 to 2024 (first discovery epoch in ZTF), and only select the targets that have at least five data points in each of the ZTF $g$- and $r$- bands. For TDEs, we further refine the samples by requiring magnitude variations in both the $g$- and $r$-bands to be greater than 0.5 mag, and excluding sources that are too close to the detection limit, i.e., those fainter than 19.5 mag at the peak. 
As a result, 50 TDEs and 7363 other types of transients are selected. The TDE sample includes 21 TDEs from the ZTF-I sample \citep{Hammerstein_2023_ZTF_I}, 16 additional TDEs from the subsequent publication \citep{tde-catalog-2}, and 13 more TDEs in TNS. More information for this sample can be found in \autoref{ZTF-TDE-score}. The constitution of the dataset is displayed in \autoref{Transient_num}.
We retrieve the light curve data of the above 7413 transients through the bulk download service of the \texttt{Lasair}\footnote{\href{https://lasair-ztf.lsst.ac.uk/}{\url{https://lasair-ztf.lsst.ac.uk/}}} API.

\begin{deluxetable*}{cccccccc}
\tablecaption{The constitutions of the ZTF light curve dataset}\label{tab:ztflc}
\tablewidth{0pt}
\tablehead{
\colhead{\textbf{TDE}} & \colhead{SN Ia} & \colhead{SN Iax} & \colhead{SN Ia(91bg-like)} & 
\colhead{SN Ib} & \colhead{SN Ic} & \colhead{SN Ib/c} & \colhead{SN Ibn}
}
\startdata
\textbf{50($0.7\%$)} & 4910($66.2\%$) & 24($0.32\%$) & 69($0.9\%$) & 160($2.1\%$) & 181($2.4\%$) & 36($0.49\%$) & 40($0.54\%$) \\
\hline
\colhead{SN II} & \colhead{SN IIP} & \colhead{SN IIn} & \colhead{SN IIb} & 
\colhead{SLSN-I} & \colhead{SLSN-II} & \colhead{AGN} & \textbf{Total} \\
\hline
1128($15.2\%$) & 151($2.0\%$) & 273($3.7\%$) & 116($1.6\%$) & 129($1.7\%$) & 77($1.0\%$) & 69($0.9\%$) & \textbf{7413} \\
\enddata
\tablecomments{With the exception of TDEs, all other data were obtained through APIs.}
\label{Transient_num}
\end{deluxetable*}

\section{Integrated Filter}\label{sec:filter}

Our filter makes use of two complementary methods, based on the parametric fitting and \texttt{Mgformer} model, respectively. We will show more details in Section \ref{sec:fit} and \ref{sec:mgformer}. Both methods possess the capability to independently identify TDE candidates and can be applied either sequentially or independently, depending on the specific needs of the analysis. Each approach offers distinct advantages, as detailed in Section~\ref{sec:CE}. The choice of implementation can be flexibly adapted based on the requirements of the task.

In addition to the light curve analysis model, we also explore the use of external data. 
These data include the angular separation between the transient source and the center of its host galaxy, the redshift and spectral classification of the host galaxy obtained from surveys such as SDSS or DESI, historical photometric baselines from time-domain surveys including ZTF and ATLAS, the mid-infrared color index $W1 - W2$.
They are only introduced after initial filtering, as we aim to minimize the dependence on external data yet they can increase the reliability of our judgments, such step are not necessary in our work.

\begin{figure*}[htbp]
\centering
\includegraphics[width=1.0\textwidth]{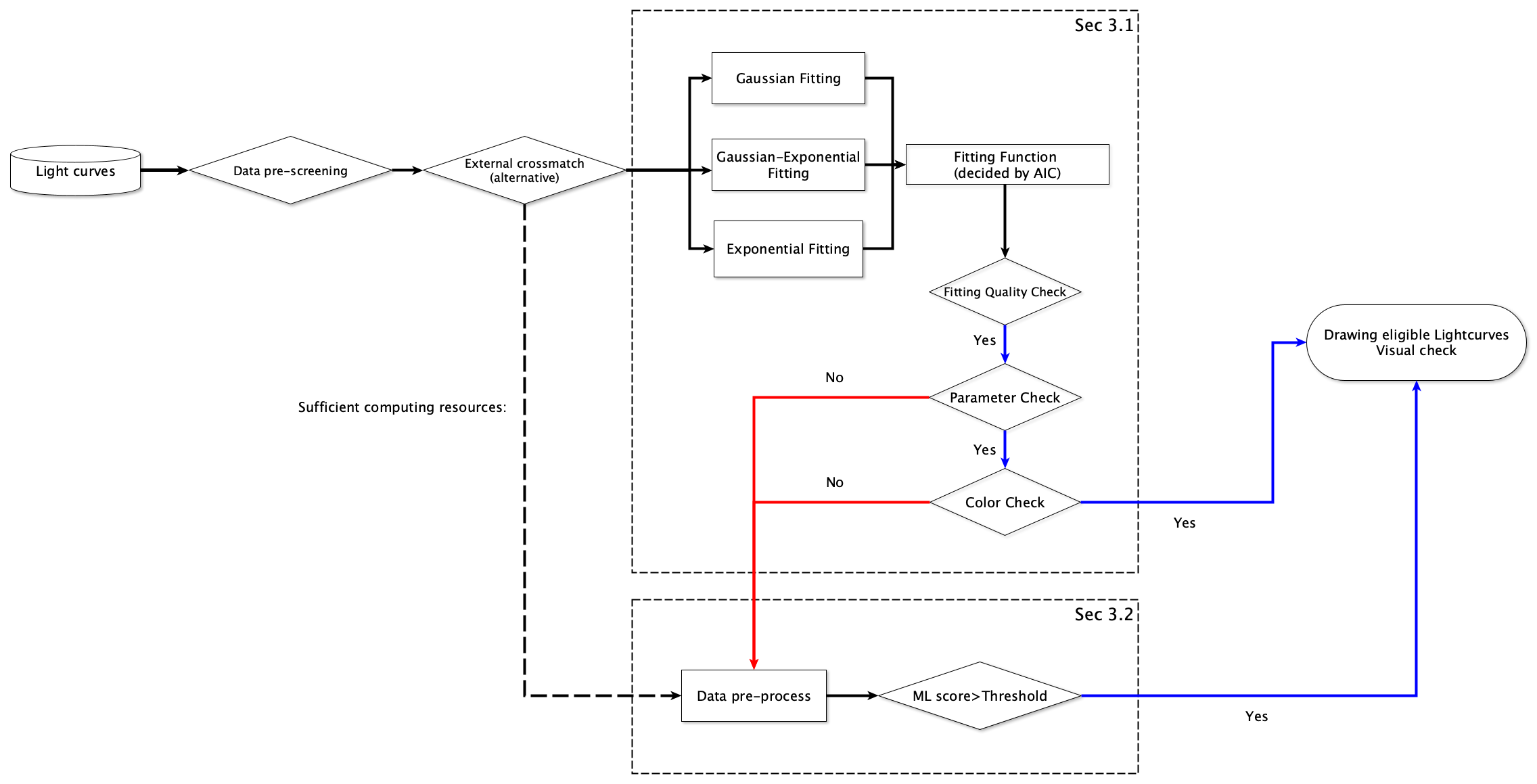}
\caption{The flowchart of the whole process. The blue arrow indicates cases where the classification result is \textbf{Y}, while the red arrow represents cases where the classification result is \textbf{N}. The dashed line indicates the scenario in which ample computational resources allow for the direct execution of machine learning classification tasks. The pre-screening of the light curve and the external crossmatch will be introduced at the beginning of the Section \ref{sec:filter}, the parameter fitting method will be introduced in Section \ref{sec:fit}, and the machine learning classification will be introduced in Section \ref{sec:mgformer}.
\label{fig:processing}}
\end{figure*}

Our entire flowchart is shown in \autoref{fig:processing}.
First, we perform a pre-screening of the data by requiring at least five data points with SNR $>$ 5 in both of $g$ and $r$ bands. Subsequently, an optional external data cross-matching can be conducted.
% , mainly to use angular separation to select sources that close to the center of its host galaxy and use color information to filter out AGN.
% , although this step is optional.  
When computational resources are sufficient, we recommend applying machine learning algorithms directly for classification (Section \ref{sec:mgformer}) to take its advantage on the early-time light curves. If insufficient, the parametric method could be used, as it is effective in excluding false light curves with much less computational resources (Section \ref{sec:fit}).
% utilizes color changes within 30 days after the peak and thus require longer-term observations (Section \ref{sec:fit}), but it is effective in excluding false light curves and requires less computational resources
% and hence is suitable for initial filtering 
% is effective in excluding false light curves 
% and hence is suitable for initial filtering under limited computational resources .
% since parametric methods are effective primarily for complete light curves and perform poorly for early-time light curves.  
% This limitation arises because color changes within 30 days after the peak require long-term observations.  
% In cases of limited computational resources, the parametric method can be prioritized for (Section \ref{sec:fit}). 
After the preliminary fitting quality check, further inspections of the fitting parameters and color evolution are performed. To avoid missing genuine candidates due to the strict partitioning of the parametric method, sources filtered out in these steps are still allowed to proceed to the machine learning classification stage.  Finally, all resulting candidates are presented on a web interface for visual inspection.

\subsection{Method based on parametric fitting}\label{sec:fit}
We employ a rise-and-fall function 
% Equation~\eqref{func_fit} 
to model the light curve:
\begin{equation}\begin{aligned}
F_{\nu}(t) =C_\nu+ F_{\nu,{\rm peak}} 
 & \times
\begin{cases}
e^{-(t-t_{\mathrm{peak}})^2/2\sigma^2}, & t\leqslant t_{\mathrm{peak}} \\
e^{-(t-t_{\mathrm{peak}})/\tau}, & t>t_{\mathrm{peak}} 
\end{cases}
\label{func_fit}
\end{aligned}\end{equation}

For the rising stage, a Gaussian function is employed, while for the descending stage, an exponential function is used. 
This function is similar as those used on ZTF TDE samples (e.g., \citealt{vanVelzen2021,Hammerstein_2023_ZTF_I}), while a baseline correction term $C_\nu$ has been added to compensate for inaccuracies in the data calibration.

\begin{deluxetable}{ccc}[htbp]
\tablecaption{Constraints for parametric fitting in each band}
\tablewidth{0pt} 
\tablehead{
\colhead{Parameter} & \colhead{Initial parameter setting} & \colhead{Range for Fitting} 
}
\startdata
$\sigma/ \mathrm{day}$ & 10 & [1, 50] \\
$\tau/ \mathrm{day}$ & 20 & [1, 1000] \\
$t_{\rm peak}/ \mathrm{day} $ & $t_{\rm max}$ & [$t_{\rm max}-20$, $t_{\rm max}+20$] \\
$F_{\rm peak}/ \mathrm{mJy}$ &$F_{\rm max}$ &[$F_{\rm max}$, $10 \times F_{\rm max}$] \\
$C/ \mathrm{mJy}$ & 0 & [$-0.1, 0.1$]\\
\enddata
\tablecomments{
% The parameter constraints applied during the fitting process.\\
$\sigma$: the rise timescale.\\$\tau$: the fading timescale.\\$t_{\rm max}$: the observation time at which $F_{\rm max}$ occurs.\\$F_{\rm max}$: the maximum observed flux in the band.}
\label{para_limit}
\end{deluxetable}

In our analysis, these baseline signals are generally attributed to residual system artifacts in the difference images, but they may also arise from imperfect calibration. The former typically appears as a small number of scattered data points preceding the main rise, whereas the latter often leads to a global offset in the differential light curve, shifting it upward or downward. This effect is present in in the light curves derived from both ZTF and WFST.

%\textbf{The reason for adopting this functional form is that it robustly captures the rising and declining timescales of the light curve with a minimal number of parameters, where the declining timescale can also be referred to as the e-folding time. Although a $t^{-5/3}$ form provides a more direct physical interpretation, the functional form
%$A \times [(t - t_p + t_{\rm scale})/t_{\rm scale}]^{-\alpha}$
%may not be preferred because the parameter $\alpha$ is extremely sensitive to the choice of $t_p$, particularly when the duration of the declining segment of the light curve is limited. Furthermore, on shorter time baselines, a degeneracy-like behavior can arise between $t_{\rm scale}$ and $\alpha$, meaning that different combinations of $t_{\rm scale}$ and $\alpha$ may produce nearly indistinguishable fits.}

We employ the $\texttt{curve\_fit}$ from the $\texttt{scipy}$ package for curve fitting. This method is based on maximum likelihood estimation and allows simultaneous fitting of multi-parameter functions. The parameter constraints and initial settings used in the fitting process are listed in \autoref{para_limit}, primarily following the approach of \citet{Hammerstein_2023_ZTF_I}. %To enhance computational efficiency,

The quality of the fit is assessed by using two parameters: the $R^2$ score and the mean absolute percentage error (MAPE), along with the time scale corresponding to the fitting parameters. The definition of MAPE and $R^2$ score are: 
\begin{equation}
\mathrm{MAPE} = \frac{1}{n}\sum_{t=1}^{n} \left|\frac{F_t - A_t}{A_t} \right|,
\label{MAPE}
\end{equation}

\begin{equation}
    R^2=1-\frac{\sum_{t=1}^n(A_t-F_t)^2}{\sum_{t=1}^n(A_t-\bar{A})^2},
    \label{R2score}
\end{equation}
where $A_t$ and $F_t$ represent the observed value and model predicted value, $\bar{A}$ represent the mean of $A_t$.
% of predicted data by fitting at the same position of the real data. 
% We calculate the absolute value of the percentage error between the data point and the fitted value at each location and average it.

For each individual source, we perform curve fitting using the Gaussian component, the exponential component, and the full piecewise function separately. The fitting is conducted independently for the $g$-band and $r$-band light curves. The optimal model is selected based on the Akaike Information Criterion (AIC), which is computed for each fit as \autoref{eq:AIC}:
\begin{equation}
    \mathrm{AIC} = 2\,k - 2\,\mathrm{ln}
    \,\hat{L}
    \label{eq:AIC}
\end{equation}
where $k$ is the number of free parameters in the model, and $\hat{L}$ is the maximum value of the likelihood function for the model. A lower AIC value indicates a better model fit, as it reflects a more favorable trade-off between goodness of fit and model complexity.

To trigger this selection, both $g$ and $r$ bands are required to have more than 5 valid detections in which signal-to-noise ratio $>$5. If so, the functions will be applied to the $g$- and $r$-band light curves to obtain the rise timescales $\sigma_g$ and $\sigma_r$, peak epochs $t_{g,\mathrm{peak}}$ and $t_{r,\mathrm{peak}}$, peak fluxes $F_{g,\mathrm{peak}}$ and $F_{r,\mathrm{peak}}$ and baseline offsets $C_g$ and $C_r$. 
We present in \autoref{fig:parameters_distribution} the distributions of the fitted parameters for all samples. 
% The shaded contour regions illustrate the parameter distribution of non-TDE sources, while red markers denote the TDE population. 
% It is evident that 
TDEs typically exhibit rising timescales $\sigma$ exceeding approximately 10 days and declining timescales $\tau$ generally greater than 20 days.

We require that $\sigma_g>10$ days and $\sigma_r>10$ days, In addition, to better limit the fitting results in both bands, we simultaneously require during the fitting process that $|\sigma_g-\sigma_r|<5$~days.

Meanwhile, both the fitting parameters and the fitting quality are used for screening. Specifically, we require the $R^2$ score for both the $g$- and $r$-bands to exceed 0.7, while the MAPE for both bands are less than 0.3. However, it should be noted that the fitting quality criteria provided here are intended primarily for reference and are most applicable in situations where a substantial fraction of bogus detections are present. For the majority of light curves that correspond to real detections, we recommend adopting a more relaxed criterion or even omitting the fitting quality check altogether.

Additional color-related screening is conducted. First, the $g-r<0$ color near the peak value is examined. Second, based on observational experience, some blue SNe near the peak exhibit rapid cooling, leading to a ``crossover" point. Specifically, when $r < g$ after the crossover, this intersection is set to occur no earlier than 20 days after the peak. Meanwhile, d$\,(g-r)/$d$\,t<0.01\  \mathrm{d}^{-1}$ in 1 month after the first observation \citep{LSST-TDE} is also checked. \autoref{fig:color_parameters_distribution} presents the peak colors of TDEs and non-TDEs, together with the color evolution within 30 days after the peak. It is evident that the vast majority of TDEs exhibit bluer peak colors and relatively slower color changes following maximum light.

Following the screening procedure based on goodness-of-fit and color variation criteria, 36 out of 50 TDEs were successfully identified, yielding a recall of 0.72. Among the remaining 7363 non-TDE sources, 53 were incorrectly classified as TDEs, resulting in a precision of 0.40. 
Without joining the Fitting Quality Check, recall will rise to 0.79, while precision will only drop to 0.3.
For definitions of recall and precision, see Section~\ref{subsec:Performance of the ML Filter}. We observed that, among the TDEs that were not successfully selected, the majority failed due to the $g$-band magnitude at peak brightness being slightly higher than that of the $r$-band, for example, ZTF22aadesap and ZTF22aacgcwv. The remaining cases exhibited relatively pronounced color evolution during the declining phase, such as ZTF20acqoiyt and ZTF20abgwfek.

\iffalse
\begin{deluxetable*}{ccccccc}

\tablecaption{Goodness of fit criterion of TDE}
\tablehead{
\colhead{Stage} &  \colhead{$R^2$} &  \colhead{$MAPE$}&  \colhead{$\sigma$}&  \colhead{$\tau$} &  \colhead{$A$}  &  \colhead{$^{*}other$} 
}

\startdata
Rising & \makecell{$R^2$(g)$>$0.7 , $R^2$(r)$>$0.7\\$R^2$(g)+$R^2$(r)$>$1.5\\$R^2$(g)$\times R^2$(r)$>$0.6} &\makecell{MAPE(g)$<$0.2\\  MAPE(r)$<$0.2 }
                & \makecell{ $\sigma$(g)$>$ 5, $\sigma$(r)$>$ 5 } & \makecell{- } & $A$(g)$>$0.5$\times$$A$(r) & \makecell{Spearman p-value(g)$<$0.05\\ 
                Spearman p-value(r)$<$0.05} \\
\cline{1-7}
Full &  \makecell{$R^2$(g)$>$0.7 , $R^2$(r)$>$0.7\\$R^2$(g)+$R^2$(r)$>$1.5\\$R^2$(g)$\times R^2$(r)$>$0.6} &\makecell{MAPE(g)$<$0.2\\  MAPE(r)$<$0.2 }
                & \makecell{ $\sigma$(g)$>$ 5\\ $\sigma$(r)$>$ 5 } & \makecell{ $\tau$(g)$>$ 10\\ $\tau$(r)$>$ 10 } & $A$(g)$>$0.5$\times$$A$(r) & - \\
\cline{1-7}
Fading & \makecell{$R^2$(g)$>$0.7 , $R^2$(r)$>$0.7\\$R^2$(g)+$R^2$(r)$>$1.5\\$R^2$(g)$\times R^2$(r)$>$0.6} &\makecell{MAPE(g)$<$0.2\\  MAPE(r)$<$0.2 }
                & \makecell{ - } & \makecell{ $\tau$(g)$>$ 10\\ $\tau$(r)$>$ 10 } & $A$(g)$>$0.5$\times$$A$(r) & -\\
\enddata
\tablecomments{The table shows the criteria used by the parameter-based fitting method to screen TDF-like light curves. The conditions marked with $\mathbf{*}$ can also be met independently.}
\label{fitting+para}
\end{deluxetable*}
\fi

\subsection{Method based on Mgformer model}\label{sec:mgformer}

\subsubsection{Data preprocessing}\label{sec:mlpreprocessing}

To facilitate model convergence and enhance generalization performance, we first apply min-max normalization to the $g$- and $r$- band flux. 
For each individual source, we identify its maximum and minimum flux values. Both photometric bands of the same source are then normalized using a common scaling factor, such that $\mathrm{flux_{max}} = \mathrm{max(flux_{g,max}}, \mathrm{flux_{r,max}})$ and $\mathrm{flux_{min}} = \mathrm{min (flux_{g,min}}, \mathrm{flux_{r,min}})$, in order to preserve the intrinsic color and color-evolution characteristics of the source.
% This normalization ensures that input values are within a consistent range, thereby stabilizing the training process and improving the model’s transferability.

To further enhance the model’s sensitivity to early-phase variability and improve its robustness, we employ a truncated data augmentation strategy. 

Before performing data augmentation, we first divide the dataset into training and test subsets to ensure that the same source does not appear in both, thereby avoiding overly optimistic evaluation results. A stratified sampling strategy is adopted, allocating 65$\%$ of each class to the training set and the remaining to the test set. After the division, data augmentation is applied independently to the training and test sets.

Specifically, each light curve is segmented at multiple quantiles of its temporal length, namely, at 0.2, 0.4, 0.6, 0.8, and 1.0 starting from the initial observation. Only segments containing more than four valid observations in both the $g$- and $r$- bands are retained to ensure data quality. Incorporating light curves of varying completeness as a form of data augmentation enables the model to learn features across different evolutionary stages of transient events. This approach improves the model’s adaptability and transferability to light curves with incomplete observations. Similar strategy has been adopted in prior work, such as  \citet{shah2025oraclerealtimehierarchicaldeeplearning}.

Following augmentation, we reconstruct each sequence to the same length, for the \texttt{Mgformer} model requires a unified length for all time sequences. For this purpose, we adopt a two-dimensional Gaussian Process (GP) interpolation with {\tt GpyTorch}~\citep{Gpytorch} to jointly reconstruct the light curves in both bands. In cases where the GP fails to converge or prone to convergence towards the prior mean due to data sparsity or noise, we instead employ a Multi-Layer Perceptron (MLP)-based interpolation. 
%While GP interpolation has become a common choice for handling irregularly sampled light curves in many works such as \citet{avocado}, it is prone to convergence towards the prior mean in data-sparse regions. This leads to undesirable flattening of the reconstructed signal. 
The probability of this phenomenon occurring is approximately 7$\%$, and in most cases, it arises from the small numerical differences between data points.

%Between the two methods, we adopt GP interpolation as the primary approach, while the MLP serves as an alternative when the GP either fails to converge or converges to a constant. This choice is motivated by the fact that MLP is more susceptible to overfitting and can introduce spurious features not present in the original data. For instance, data points with significant deviations from the light curve may induce local abnormal oscillations in MLP, whereas the GP, due to its smoothing prior and kernel function, is more robust to a few outliers and better preserves the overall shape of the light curve. Although both methods have their advantages, the interpolation obtained via the GP is generally smoother.

We adopt GP interpolation as the primary approach because it provides smoother results and is generally more robust to a small number of outliers, thanks to its smoothing priors and kernel-based structure.  In contrast, MLPs are more prone to overfitting. %and may introduce artificial features that are not present in the original data. 
In cases where the GP fails to converge or converges to a constant function, we instead choose the MLP, since it remains capable of producing an interpolation under such circumstances. %For example, data points with large deviations from the main light-curve trend can cause local oscillatory artifacts in the MLP, whereas the GP better preserves the global structure.

In the end, we apply min-max normalization again.

\begin{deluxetable}{lrrr}[htbp]
\tablecaption{Samples in the training and test sets.\label{tab:data_distribution}}
\tablehead{
\colhead{\textbf{Class\ \ }} & 
\colhead{\textbf{\ \ Training Set}} & 
\colhead{\textbf{\ \ Test Set}} & 
\colhead{\textbf{\ \ Total}}
}
\startdata
TDE         & 142  & 77  & 219  \\
SN~Ia       & 11267 & 6086 & 17353 \\
SN~Ib/c     & 725  & 385  & 1110 \\
SN~II       & 3906 & 2106 & 6012 \\
SLSN        & 486  & 283   & 769  \\
AGN         & 147  & 73  & 220 \\
\hline
\textbf{Total} & \textbf{16673} & \textbf{9010} & \textbf{25683} \\
\enddata
\tablecomments{Sample counts of each transient class used for training and testing the classifier after data augmentation.}
\end{deluxetable}

\subsubsection{Performance of the ML Filter} \label{subsec:Performance of the ML Filter}

Given the limited number of available TDE samples, it is crucial to maximize data utilization. Therefore, after obtaining promising results in test data, the previously designated test set is incorporated into the training set to enhance the model's learning capacity. In this section, we only present the performance on the test set.

The hyperparameters set by the training the model and the training epoch are detailed in the \autoref{hyperparameter}. 
%The hyperparameters are configured with values smaller than the model's default settings \citep{MgFormer}, 
%as increasing the model complexity beyond this baseline does not lead to significant performance improvements. 
Based on our experiments, the hyperparameters listed in the table represent the minimum values required to ensure model convergence. We also explored the default settings from \citep{MgFormer} and larger parameter settings, including various parameter combinations, and found no significant difference in the convergence outcomes. However, excessively large parameter values substantially increase the training time.

This model was deployed for training and testing on a single NVIDIA A100-PCIE-40GB. Each training epoch takes about few minutes.
%The model obtained after training is called \textbf{Model 1}.
\begin{deluxetable}{c|c}[htb]
% \tablewidth{\textwidth}
\tablecaption{Hyperparameters set for training}
\tablehead{
\colhead{  Hyperparameters  } &  \colhead{\ \ Value\ \ }
}
\startdata
multi$\_$group & [1,2]\\
batch & 1\\
lr &\ \ 0.0001\ \ \\
nlayers & 2\\
emb$\_$size &32\\
nhead & 4\\
emb$\_$size$\_$c & 32\\
masking$\_$ratio & 0.15\\
\ \ \ \ ratio$\_$highest$\_$attention\ \ \ \ \ & 0.35\\
dropout& 0.01\\
nhid& 32\\
nhid$\_$c& 32\\
Training epochs & 40
\enddata
\tablecomments{The definition and default setting of these hyperparameters are introduced in \citet{MgFormer}.}
\label{hyperparameter}
\end{deluxetable}

Our evaluation is based on three key metrics: precision, recall and F1-score. Precision quantifies the accuracy of TDE detection, while recall measures the completeness of TDE identification, The F1-score is the harmonic mean of precision and recall. 
% A higher F1-score indicates a more robust and balanced classification performance of the model. 
The calculation of these metrics requires a four-fold classification, namely
% For classification, we categorize results into four groups:
\begin{tight_enumerate}
    \item \textbf{True Positive (TP)}: TDE correctly classified as TDE,
    \item \textbf{False Positive (FP)}: Non-TDE incorrectly classified as TDE,
    \item \textbf{False Negative (FN)}: TDE incorrectly classified as non-TDE,
    \item \textbf{True Negative (TN)}: Non-TDE correctly classified as non-TDE.
\end{tight_enumerate}

Precision, recall and F1-score are defined as
% follows in Equation~\eqref{precision_recall}:
\begin{equation}
\mathrm{Precision} = \frac{\mathrm{TP}}{\mathrm{TP} + \mathrm{FP}},
\end{equation}
\begin{equation}
\ \ \ \,\,\mathrm{Recall} = \frac{\mathrm{TP}}{\mathrm{TP} + \mathrm{FN}},
\end{equation}
\begin{equation}
\mathrm{F1\text{-}score} = \frac{2 \cdot \mathrm{Precision} \cdot \mathrm{Recall}}{\mathrm{Precision} + \mathrm{Recall}}. 
\end{equation}

We adopt the \texttt{Adam} optimizer \citep{Adam} and evaluate the model performance after each epoch using two metrics: the F1 score for TDE classification on the test set ($\mathrm{F1_{TDE}}$) and the global F1 score across all classes ($\mathrm{F1_{{total}}}$). The epoch yielding the maximum sum of $\mathrm{F1_{TDE}} + \mathrm{F1_{total}}$ is selected as the final model checkpoint.
Despite the substantial class imbalance in the dataset, we employed the standard (unweighted) cross-entropy loss and obtained satisfactory performance, which we attribute primarily to the effectiveness of our data augmentation strategy.

\begin{figure*}[htbp]
\centering
\includegraphics[width=0.9\textwidth]{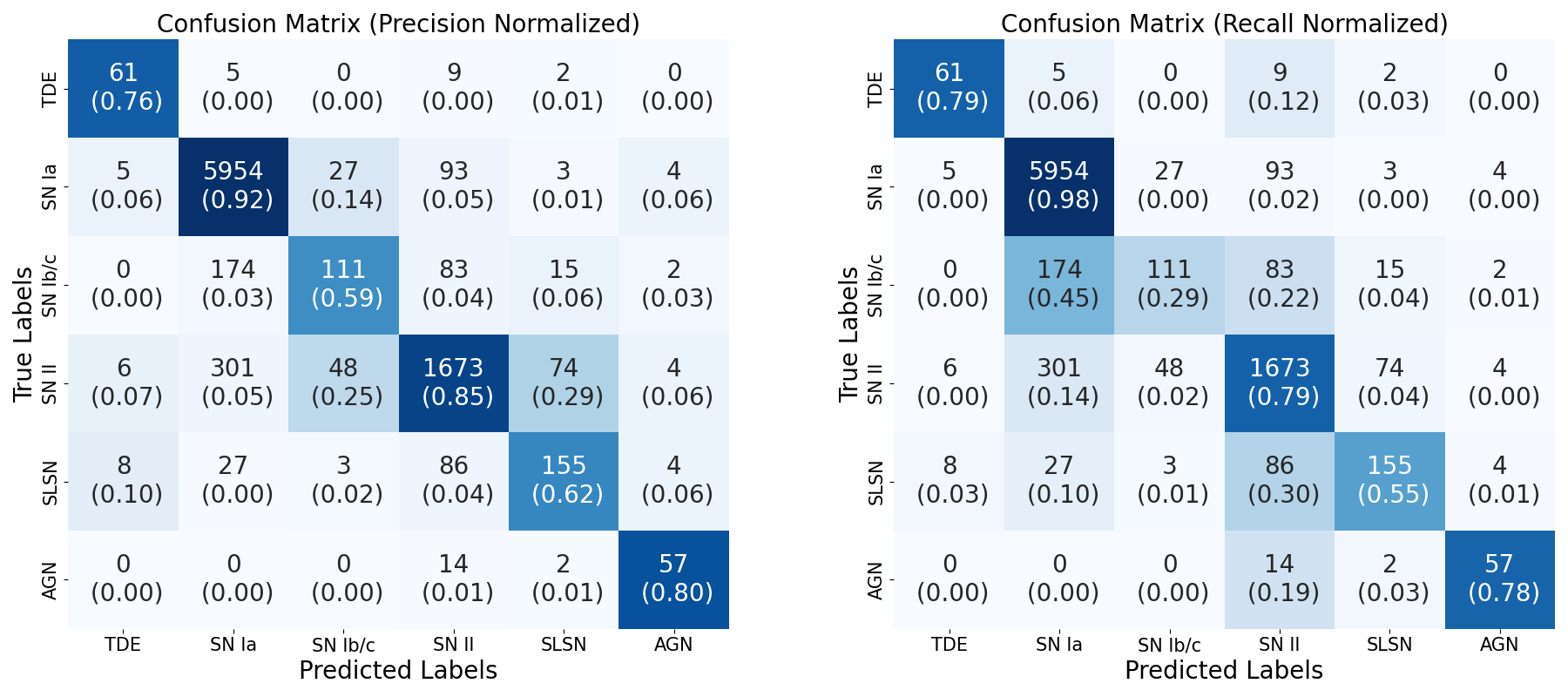}
\caption{%The performance on the test set is evaluated using a strictly random partitioning of the data into training and test sets. Without fine-tuning the decision boundary, the classification achieves a precision of 0.76 and a recall of 0.79. Notably, the classification process relies solely on the information provided by the light curve.
The performance on the test set, the classification achieves a precision of 0.76 and a recall of 0.79. Notably, the classification process relies solely on the information provided by the light curve.
\label{fig:TDE martix}}
\end{figure*}

\begin{figure*}[htbp] 
\centering
\includegraphics[width=0.3\textwidth]{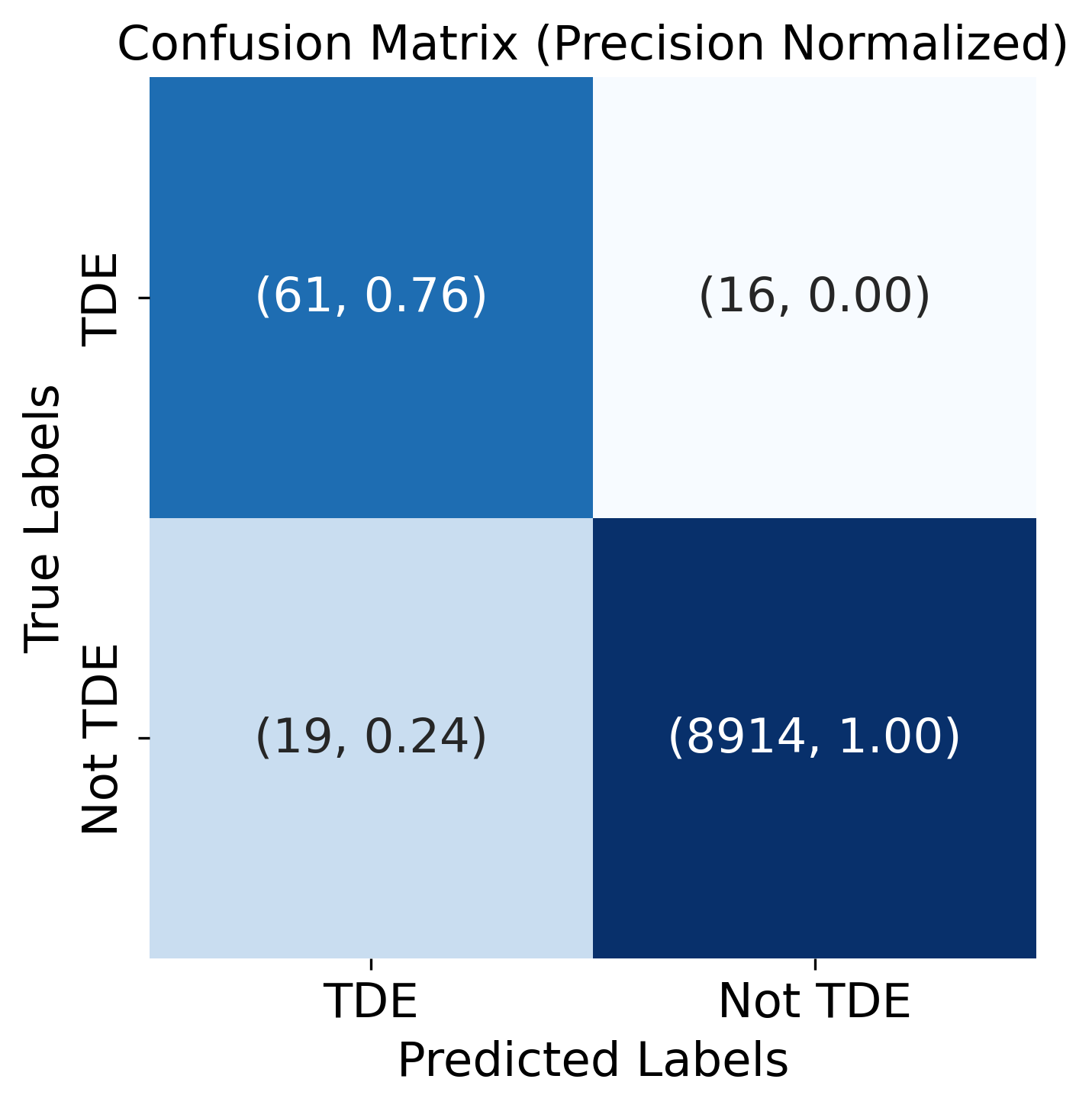}
\hspace{2cm}
\includegraphics[width=0.305\textwidth]{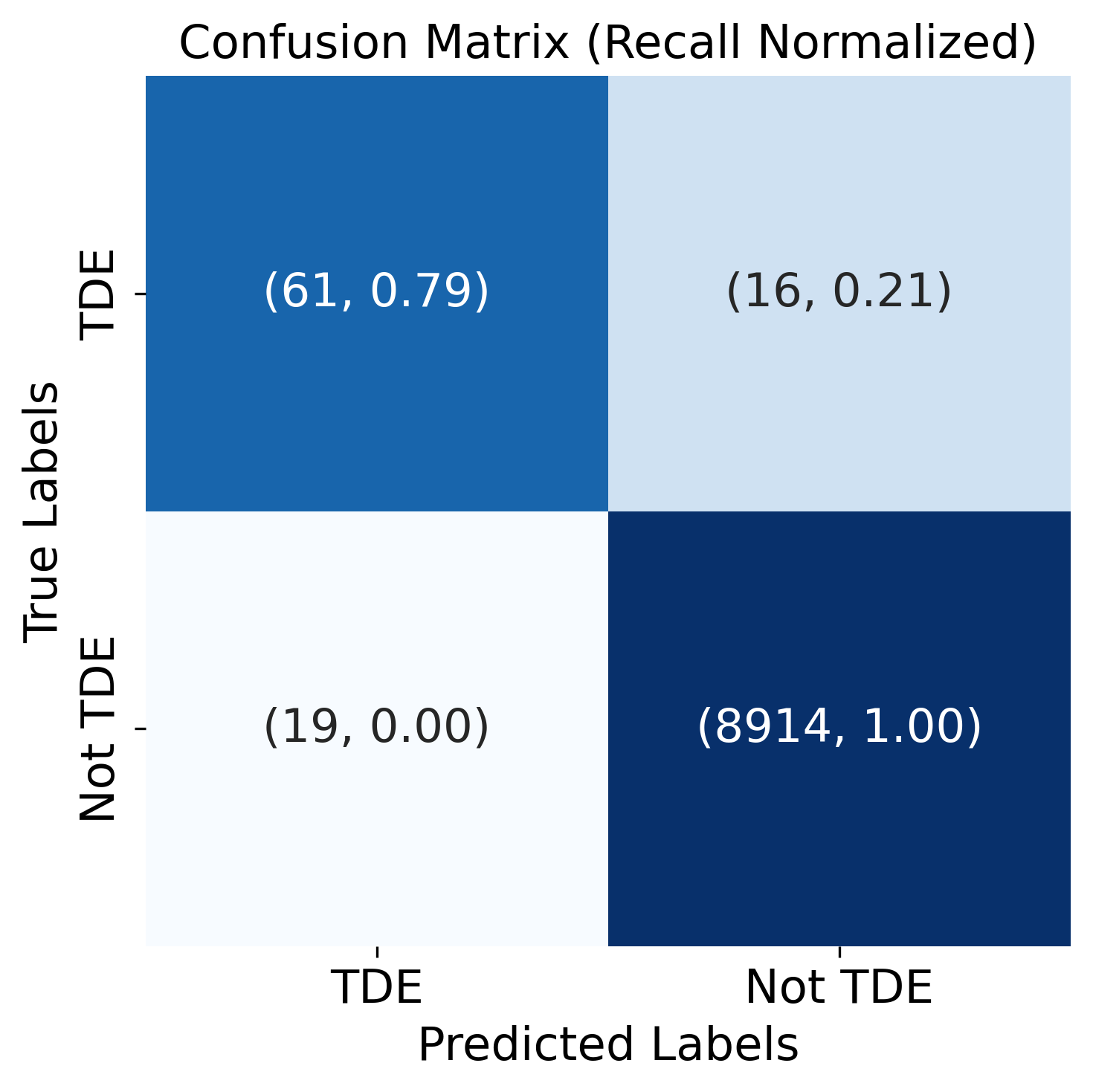}
\caption{The confusion matrix shown here is analogous to that in \autoref{fig:TDE martix}, with the distinction that all non-TDE classes have been merged into a single ``Not TDE” category.
\label{fig:2x2}} 
\end{figure*}

The evaluated performance of the ML filter
% effectiveness of the algorithm
% we trained 
is demonstrated as follows. We present the confusion matrix on the test set in \autoref{fig:TDE martix}. This model achieves a precision of 0.76 and a recall of 0.79 on the test set. For simplicity, the weights produced by the classification layer of the machine learning algorithm are hereafter referred to as the ``TDE score" throughout this study. Given our primary interest in TDE classification, we represent the classification performance for TDEs using a simplified 2$\times$2 confusion matrix which showed in \autoref{fig:2x2}.

\begin{figure*}[htbp]
\centering
\includegraphics[width=0.7\textwidth ]{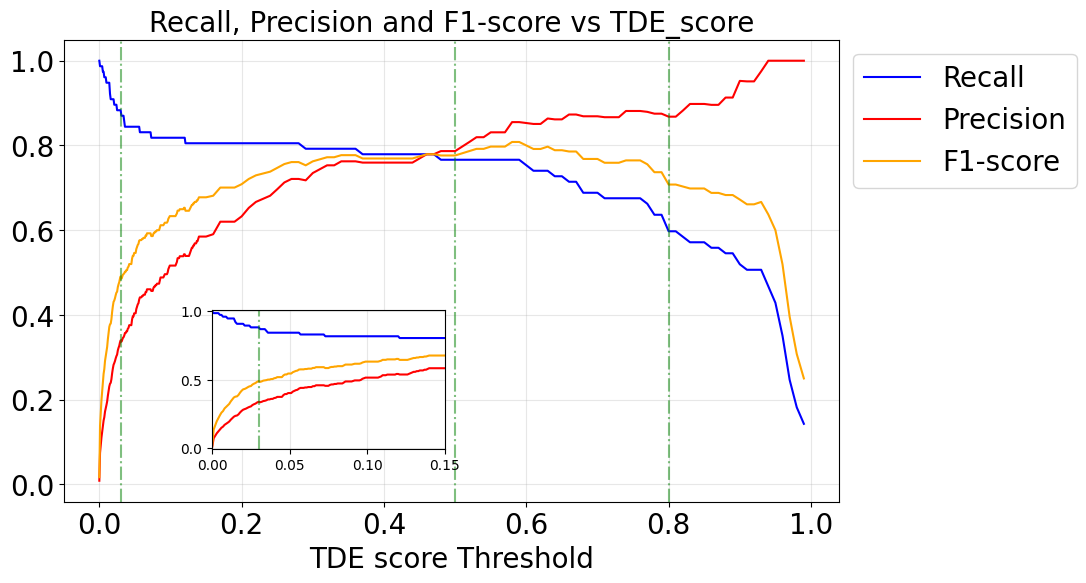}
\caption{The relationship between the TDE score and precision, recall and F1-score is depicted, with the blue solid line representing recall and the red solid line representing precision. The intervals (0, 0.15) is magnified in the left subplots. The three vertical green dash-dot lines represent the thresholds of the three strategies.
\label{fig:TDE mart 2}}
\end{figure*}

We further explore the relationship between the TDE score threshold and TDE recall, precision and F1-score in \autoref{fig:TDE mart 2}. 
Based on this relationship, we assign three strategies based on TDE score thresholds: 0.03, 0.5 and 0.8, that yield recall$\approx$0.9 (recall-oriented strategy), precision$\approx$0.9 (precision-oriented strategy) and recall$\approx$precision (balanced strategy), respectively.

\subsection{Application of the integrated filter on latest ZTF data}

The final routine of our filter is described as follows. First, the parametric fitting method is employed to evaluate the shape and color trends of the light curves. Subsequently, the ML model is utilized to assign scores. In the \autoref{ZTF-TDE-score}, we present the scores for TDEs included in the training set.

To assess the reliability of our model on new data, we retrieved the latest light curves from the ZTF by using the Lasair API. The selection criteria for light curves are
% were intentionally kept 
straightforward: (1) the source must have been observed at least once within 10 days prior to the date of our test, April 28, 2025, and (2) the light curve must contain more than 10 data points. No constraints were imposed regarding angular separation from the host galaxy center or the host galaxy properties. (3) The source's name must begin with "ZTF25", which means this source is officially discovered in 2025.

Applying these criteria, we collected 129 light curves in total.
% , including three associated with spectroscopically confirmed TDEs. 
Following the methodology described in previous sections, we selected candidates based on a cut of TDE score$>$0.5.  
This selection cut finds all 4 spectroscopically confirmed TDEs with 100\% precision. The names, discovery dates, and
TDE scores of these sources are summarized in \autoref{table:ZTF_candidate_new}. In particular, ZTF25aajjeon was 
% selected as a highly promising candidate and 
% classified as a featureless TDE at $z=0.6278$ 
selected as a highly promising candidate  4 days before it is classified as a featureless TDE at $z=0.6278$ \footnote{\href{https://www.wis-tns.org/astronotes/astronote/2025-131}{\url{https://www.wis-tns.org/astronotes/astronote/2025-131}}}, demonstrating the capability of finding distant TDEs. 

\begin{deluxetable*}{ccccc}[t]

\tablecaption{ZTF TDE Candidates Selected by Model}
\tablehead{
\colhead{ZTF Source} & \colhead{IAU Name} & \colhead{Transient Type (TNS)} & \colhead{Discovery Date} & \colhead{TDE Score}
}

\startdata
ZTF25aafofcs & TDE 2025bri & TDE-He & 2025-02-15 & 0.59 \\
ZTF25aafwfzz & TDE 2025cyh & TDE & 2025-02-15 & 0.86 \\
ZTF25aagevje & TDE 2025chm & TDE & 2025-02-20 & 0.94\\
ZTF25aajjeon & TDE 2025hbw & TDE & 2025-03-19 & 0.98\\
\enddata
\tablecomments{This table shows TDE candidates that we select from 129 transients that named after ``ZTF25'' and have at least one detection within 10 days before April 28, 2025, the date we select.
% and the search was carried out on April 28, 2025, while the 
ZTF25aajjeon is classified on May 2, 2025.}
\label{table:ZTF_candidate_new}
\end{deluxetable*}

\section{Application on WFST Survey data}\label{sec:supply-real-data}

The sky survey strategy of WFST consists of two main components: the Wide-Field Survey (WFS) and the Deep High-Cadence $u$-band Survey (DHS) program \citep{wfst-sci}. The cadence of these surveys is designed to align with their respective scientific objectives, with each occupying approximately 45\% of the total observation time. During the operation of WFST, we generate a vast dataset. The raw data are processed using a standardized data pipeline \citep{WFST-pipeline}. 

The WFS program will cover $\sim$8000 deg$^2$ of the northern sky, utilizing $u, g, r, i$ bands with a single-exposure time of 30 s. % All $u$-band observations will be conducted during dark or gray nights, and whenever possible, observations will be obtained in at least two bands per night. 
In addition to WFS, the DHS program is designed to leverage WFST’s superior $u$-band imaging performance for time-domain surveys. DHS will conduct regular monitoring of two $\sim360$ deg$^2$ sky regions - ``Spring" and ``Autumn" fields near the celestial equator. %In this study, we present two representative examples selected for presentation. 
Following the completion of reference image preparation, the DHS of WFST has been systematically conducting planned science survey since March 2024. Reference images for the WFS are currently under preparation, with construction having begun in late 2024. %The stage of real-time survey operations and image subtraction is expected to begin in the coming months. 
%WF-difference image have been started load since 2025-08-15
In the meantime, we have initially applied our method to the deep-field component of the survey.

%The WFST pilot survey commenced in March 2024 and continue until July 2024, 

%The pipeline consists of several key steps: first, for each single-frame image obtained by the CCD, instrumental noise is subtracted, followed by image characterization and image calibration. To derive the corresponding data products, single-frame images are subsequently integrated. This process includes retrieving template images, transient detection, and ultimately, alert distribution. The most crucial differential images for our work are produced during the transient detection step, while source association occurs in the alert distribution phase. The construction of our light curves relies on alert packages generated from the differential images and source association.

We apply the same min-max normalization to the WFST light curves, followed by a standardization step using the mean and variance derived from the training set. This ensures that the input scale remains consistent with that of the training samples, thereby maintaining compatibility and stability during inference.

Since the ZTF dataset used for training contains only the $g$ and $r$ bands, our classifier is currently limited to using these two bands when applied to WFST data. In the future, as WFST observations accumulate, we plan to retrain and evaluate the model using the WFST dataset itself, which includes additional photometric bands. We expect that incorporating more bands will further improve the classification performance by providing richer color information.

%========================

\subsection{WFST TDE candidates}

During each quarter of the deep field survey, WFST is capable of finding $\sim$8 TDEs 
\citep{wfst-tde}. We have found about 20 candidates by the method mentioned in Section~\ref{sec:filter} till May 1, 2025 in total, but unfortunately all of them have been fainter than 21 mag in both $g$ and $r$ band since their discoveries, which is too faint for our spectroscopic follow-up observations. We expect the situation would be fundamentally improved, as the WFS differential database will be ready in the next couple of months and the anticipated discovery rate will increase by an order of magnitude. 

For reference, we hereby present two promising TDE candidates selected by our method: 
WFST J134750.34+025201.4 and WFST J015306.85+065606.7 
(alert name: WFST0467evz).
\footnote{{{After the formal survey begins, all sources are assigned a unique alert name in the format WFST(Date since first WFST exposure)+abc or WFST-PS+YYMMDDabc(where PS denotes ``Pilot Survey"). However not all source has it own alert name, such as the first source. In parallel, an internal naming convention using the format WFST+J2000 is applied to all sources, based on their celestial coordinates. Beginning July 1, 2025, the official WFST alert naming system will be adopted. The standardized format will be WFSTYYMMDDabc, where YYMMDD indicates the discovery date, and abc represents the order of detection on that day. The suffix abc follows a based 26 sequence: a$\sim$z, followed by aa$\sim$zz, in a manner consistent with the TNS naming convention.}}}

%Regarding the naming convention of WFST sources: During the pilot survey phase, the naming rules were not yet standardized. Only sources that were confirmed for submission to the TNS were manually assigned names. As a result, the first source mentioned in this work does not have an alert name, whereas the supernova discussed in the appendix follows the format WFST-PS+YYMMDDabc, where PS denotes ``Pilot Survey".

%After the formal survey begins, all sources are assigned a unique alert name in the format WFST(Date since first WFST exposure)+abc. In parallel, an internal naming convention using the format WFST+J2000 is applied to all sources, based on their celestial coordinates.

%Beginning July 1, 2025, the official WFST alert naming system will be adopted. The standardized format will be WFSTYYMMDDabc, where YYMMDD indicates the discovery date, and abc represents the order of detection on that day. The suffix abc follows a based 26 sequence: a$\sim$z, followed by aa$\sim$zz, in a manner consistent with the TNS naming convention.

% However, these candidates were not included in spectral follow-up observations due to their faintness and the lack of available spectral resources. While the absence of spectroscopic confirmation undermines their credibility and prevents multi-wavelength verification, we believe that these two candidates highlight the potential of the Wide-Field Survey Telescope (WFST) in discovering TDEs in very faint galaxies.
For the former source, we demonstrate the potential of WFST to detect faint TDEs. The latter source is currently among the brightest WFST TDE candidates, featuring the best $u$-band sampling and a relatively high classification score.

\begin{figure*} 
\centering
\includegraphics[width=0.9\textwidth]{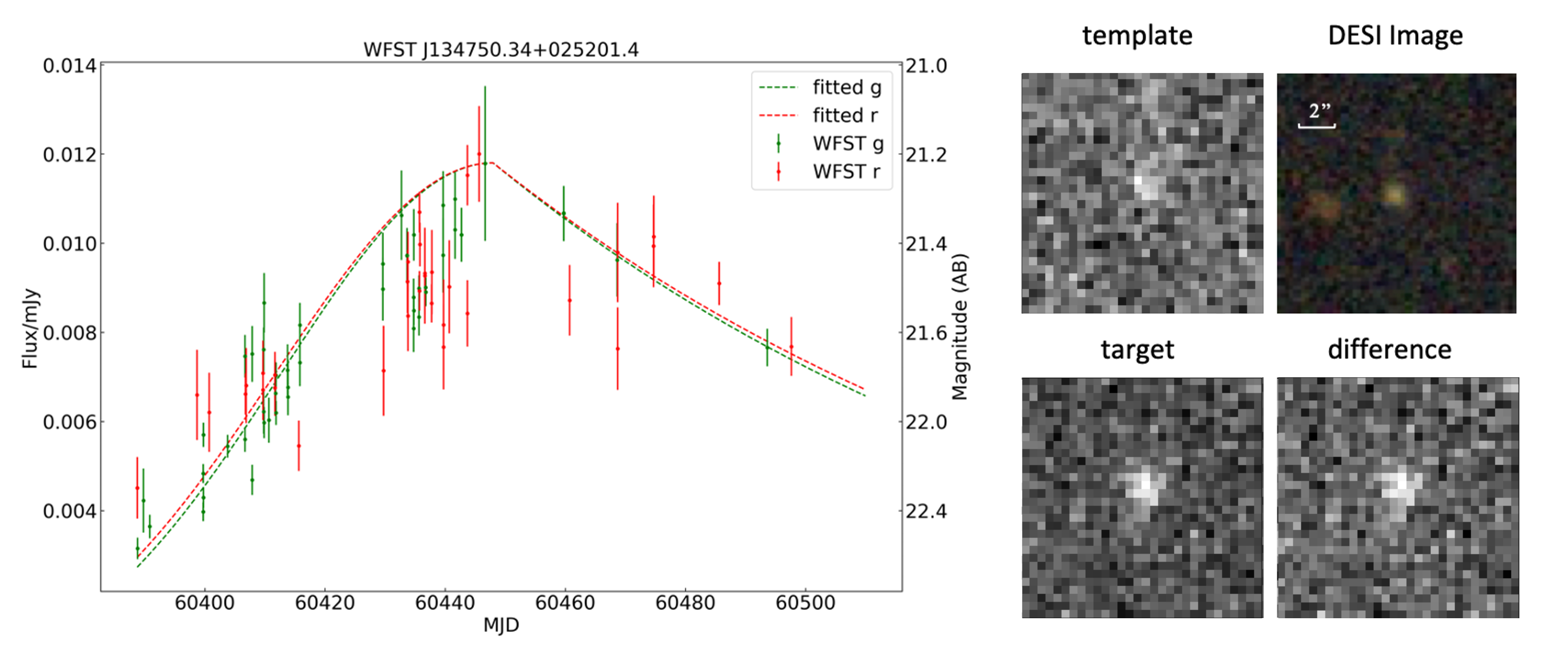}
\caption{\textbf{Left panel:} The differential light curve of WFST J134750.34+025201.4. The green and red
data points represent the $g$- and $r$-band differential flux respectively, while the green and red dotted lines represent the best fits to the rise-and-fall function (Equation \ref{func_fit}). \textbf{Right panel:} WFST $r$-band cutouts of WFST J134750.34+025201.4 on MJD = 60444: ``target", ``template" and ``difference'' mark the reference, science and difference cutouts, respectively.
% They are extracted at  May 14, 2024 with $r$-band.
% ``difference'' is the difference image, and 
The cutout size is $16.5^{\prime\prime} \times 16.5^{\prime\prime}$. 
% It can be seen that the host galaxy is too faint. 
The upper right corner shows the $griz$-colored cutout 
% image near RA=206.9597, DEC=2.8671 
in DESI Legacy Surveys DR10.
% (\url{https://www.legacysurvey.org/dr10/}).
\label{fig:candidate_lc_1}}
\end{figure*}

\begin{figure*}
\centering
\includegraphics[width=0.9\textwidth]{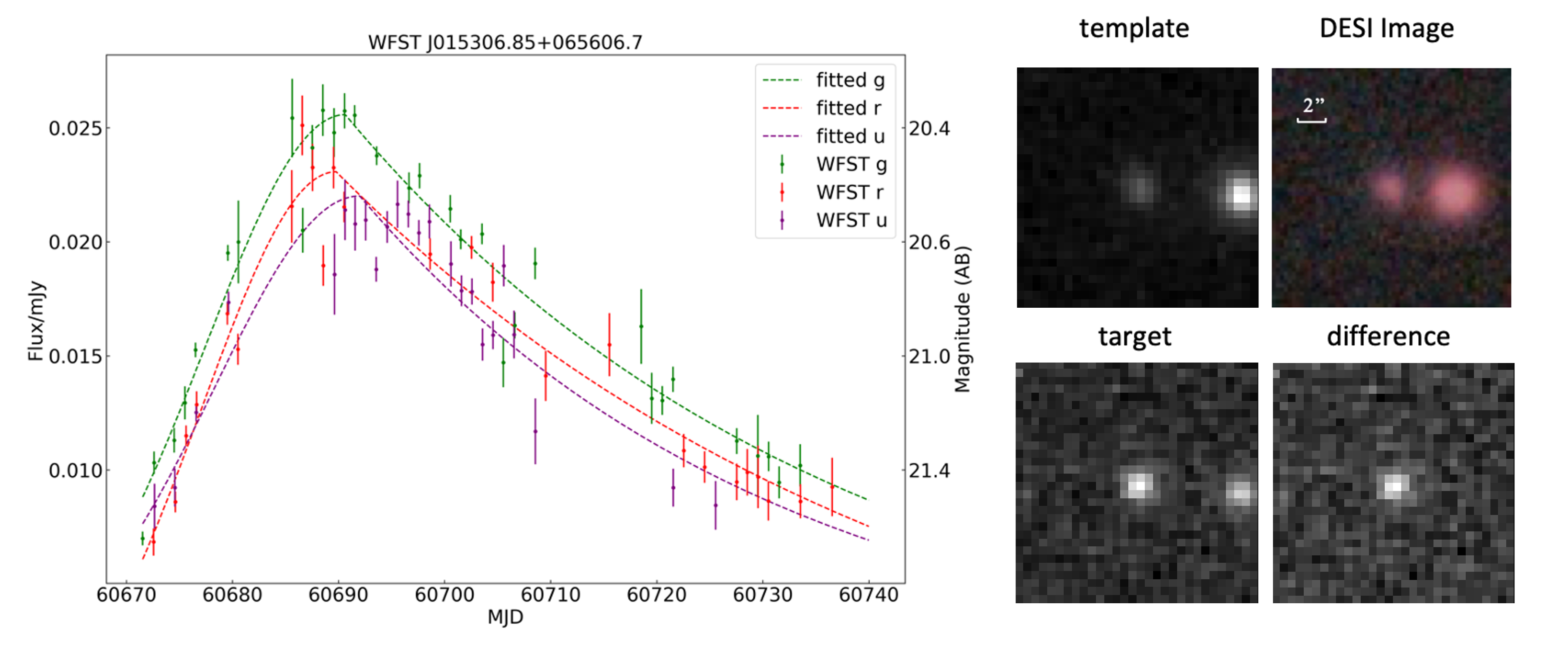}
\caption{Same as \autoref{fig:candidate_lc_1}, but for WFST J015306.85+065606.7. It has a well-sampled $u$-band light curve. The WFST cutout images were taken in the $r$ band and created on MJD = 60686.
% (RA = 28.2785, DEC = 6.9352)
\label{fig:candidate_lc_2}}
\end{figure*}

WFST J134750.34+025201.4 (TDE score = 0.0592), is associated with a host galaxy in the HSC-Wide catalog \citep{hsc}, with a photometric redshift of $z_{\text{phot}} = 0.840^{+0.093}_{-0.350}$ \citep{HSC_redshift} within 0.082 arcsec. Its $g$, $r$, $i$, and $z$-band magnitudes, as recorded in the DESI Legacy Imaging Surveys \citep{DESILS} DR10 catalog, are 23.64$\pm$0.20, 22.97$\pm$0.16, 22.34$\pm$0.09 and 21.97$\pm$0.14, respectively. The observed rise timescale and color evolution near the peak are consistent with a TDE classification. However, we cannot entirely rule out alternative explanations, such as a SLSN
% , \citealt{SLSN}
or a flare from an AGN
% , \citealt{CLAGN}
based solely on the available light curves.
Because some SLSNe may also exhibit bluer colors and slower color evolution \citep{ZTF-SLSN}%(e.g., SN 2015bn; \citealt{Nicholl2016})
, the possibility of such confusion cannot be fully excluded. Moreover, the possibility of an AGN flare cannot be entirely excluded as alternative interpretations, for the lack of information to confirm whether there is AGN activity in this source. Relying solely on the light curve in the $g$- and $r$-bands is insufficient to conclusively rule out the presence of an AGN, as AGN activity can exhibit similar photometric variability \citep{vanVelzen2021}. %for transients whose classifications remain ambiguous such as PS1-10adi \citep{Kankare2017,Jiang2019} and ASASSN-18jd \citep{Neustadt2020}. 
Additional multi-wavelength data are needed to enable a more definitive interpretation. The absence of pre-outburst photometric coverage further complicates the identification and characterization of the transient's origin.

The source's light curve, along with the reference image at the peak position (template), the scientific image (target), the difference image, and the host galaxy image, are comprehensively presented in \autoref{fig:candidate_lc_1}.

WFST J015306.85+065606.7 (TDE score = 0.9135), is 
% not detected in the WFST pilot survey but is instead 
identified during the observations in late December 2024. Since its first detection on December 27, 2024 (MJD = 60671), it has been well monitored by WFST until March 2, 2025 (MJD = 60736). 
% The duration of over 2 months is significantly longer than typical SN light curves, making this candidate more convincing than the previous source. 
% In addition to its rise timescale and color evolution near the peak, this source benefits from relatively complete $u$-band monitoring, which provides further insight into its nature. Throughout its evolution, 
Its rise and fall timescale and constant $g-r$ color of $<$ 0 are well consistent with TDEs. In addition, the well-sampled $u$-band light curve of WFST shows an evolution trend that synchronous with $g$ and $r$ bands. If the $u$-band flux is slightly underestimated, the corrected color will be typical for a TDE. 
%With the help of the relatively complete detections in the $u$-band, which enables further validation through its $u$-band light curve, we find that this object shows almost no significant color change from the decline phase up to approximately 60 days after peak brightness. This behavior is uncommon for SN, which typically exhibit low $u$-band flux at early times and a more rapid decline as the ejecta cools. Based on these features, the probability of this transient being a SN appears relatively low. Nevertheless, due to the limited availability of supporting data, we cannot completely rule out the alternative possibilities discussed above.
The $g$, $r$, and $z$-band magnitudes of its host galaxy in the DESI Legacy Imaging Surveys DR10 catalog are 22.20$\pm$0.09, 20.91$\pm$0.04, and 20.09$\pm$0.05, respectively. A similar image as that of WFST J134750.34+025201.4 is presented in \autoref{fig:candidate_lc_2}.

\iffalse
\begin{figure*}[hb!]
\centering
\includegraphics[width=0.7\textwidth]{light_curve_plot_2.png}
\caption{Same as \autoref{fig:candidate_lc_1}, but for WFST J015306.85+065606.7 (RA = 28.2785, DEC = 6.9352). This source has a relatively complete $u$-band light curve.
\label{fig:candidate_lc_2}}
\end{figure*}

\begin{figure*}[hb!]
\centering
\includegraphics[width=0.7\textwidth]{host2.png}
\caption{Same as \autoref{fig:candidate_1_host}, but for WFST J015306.85+065606.7 (RA = 28.2785, DEC = 6.9352).
\label{fig:candidate_2_host}}
\end{figure*}
\fi

Despite the unfortunate miss of spectroscopic confirmation, these two candidates demonstrate the capabilities of WFST in detecting potential TDEs in faint and distant galaxies through our TDE classifier.

\section{Discussion} \label{sec:discussion}

%During the three-month WFST pilot survey, we identified 20 TDE candidates using the method described earlier. The scores assigned to these candidates are presented in Table X, with their corresponding light curves provided in the appendix. For clarity and conciseness, only the most promising candidates are included in this report.

\subsection{Ability of capturing early TDE candidates}

\begin{figure}[!htbp] 
\includegraphics[width=0.9\columnwidth]{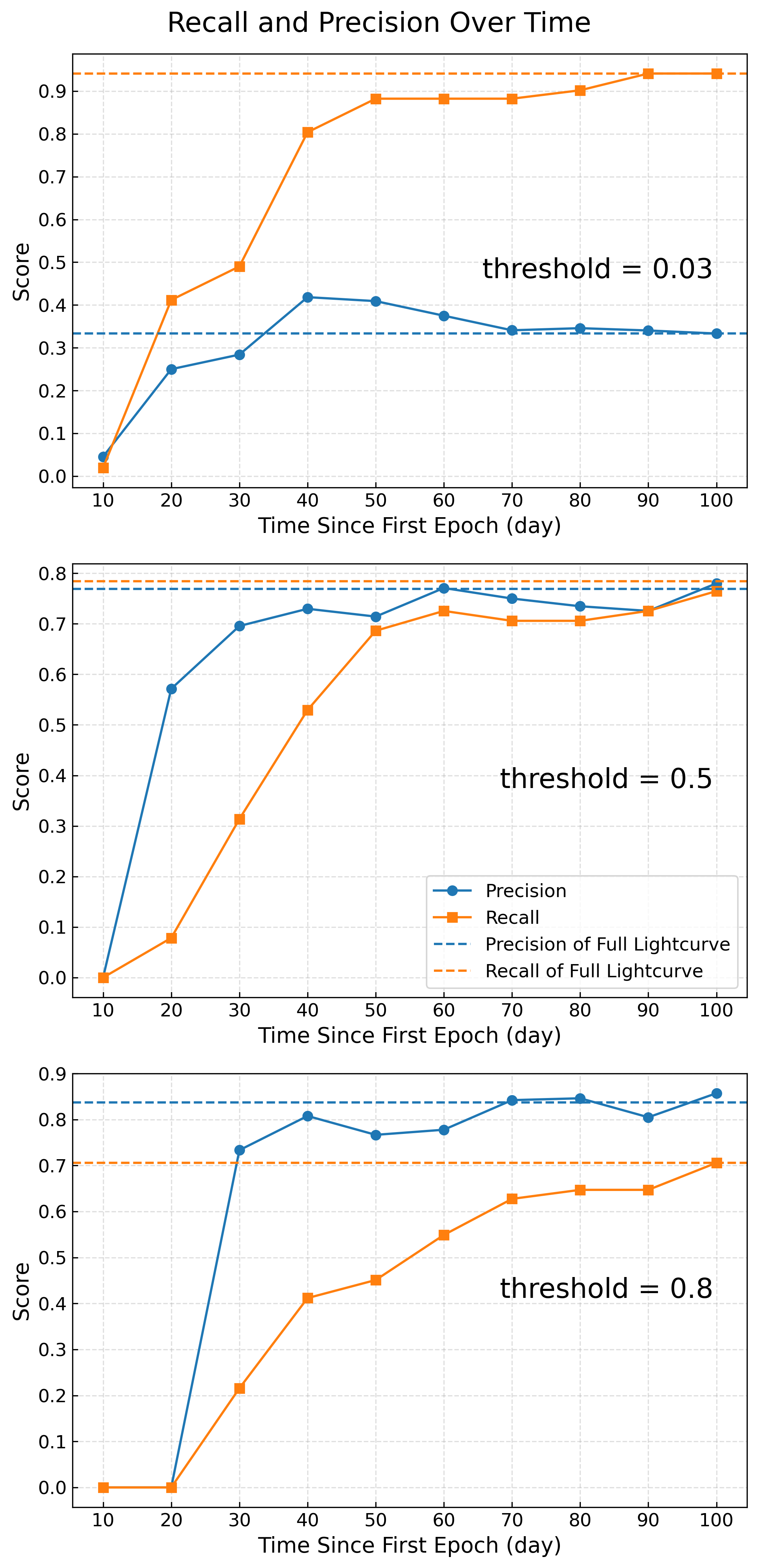} 
\caption{Recall and precision as a function of time since the first observational epoch which corresponding to the three strategies described in Section~\ref{subsec:Performance of the ML Filter}. 
%The label ``Inf'' on the x-axis indicates the result obtained using the complete light curve.
\label{fig:time_vs_recall_precision}} 
\end{figure}

As introduced in Section \ref{sec:mlpreprocessing}, data augmentation has been employed to enhance the capability in handling early light curves.
% To evaluate the model's performance on incomplete early-time data, we employed data augmentation techniques with the specific goal of enhancing its capability in handling early light curves. While the global classification performance is illustrated in \autoref{fig:TDE martix}, this representation reflects the model's overall behavior on the augmented dataset rather than its sensitivity to varying data completeness.
To more rigorously assess the model's classification performance across different phases of transient evolution, we hereby evaluate the recall and precision as functions of time since the first epoch. Based on the results shown in \autoref{fig:TDE mart 2} and strategies mentioned in Section~\ref{subsec:Performance of the ML Filter}, we calculate these metrics at 10-day intervals, plotting them against the number of days elapsed since the initial observation, the results are presented in \autoref{fig:time_vs_recall_precision}.

Specifically, the balanced strategy (TDE score threshold = 0.5) 
% offers a well-balanced trade-off between precision and recall across different stages, and 
works out best in finding TDEs at early phases, as it reaches a precision of 0.57 using the light curves within first 20 days, and the recall quickly rises given longer light curves. Regarding the deeper flux limit of WFST, this strategy should enable WFST to discover TDEs earlier than ZTF. In comparison, the precision-oriented strategy (threshold = 0.8) and the recall-oriented strategy (threshold = 0.03) are more suitable for the cases of limited and abundant follow-up resources, respectively.

\subsection{Trade-off between performance and time}
% Computational efficiency}
\label{sec:CE}

The parametric fitting and {\tt Mgformer}-based method are both able to handle the classification task independently. A brief comparison of the strengths and limitations of these two methods is provided in Table \ref{table:fit_vs_ml}. 

In details, regarding the individual performance, the {\tt Mgformer}-based method is superior in metrics (recall: 0.79 vs. 0.72; precision: 0.76 vs. 0.40), flexibility (assign scores that allow for different strategies) and versatility (capability of multi-class classification, e.g., SNe, introduced in Section \ref{sec:classifySN}).

Despite the performance, the parametric fitting method should consume shorter time as it only requires one curve fitting step. To quantify the speed, we run a test 
% conduct a time consuming estimation for the above process by running 
single-threaded on a personal computer (Apple M2 Pro, 16 GB RAM). This test uses the same ZTF light curve dataset introduced in Section \ref{sec:data_collect} and Table \ref{tab:ztflc}.
% of 7413 light curves, which were downloaded from \texttt{Lasair}. 

In the parametric fitting stage, the total processing time for the test is 179~s, yielding an average processing time of 0.024~s per source. However, the subsequent light curve reconstruction and {\tt Mgformer}-based classification cost a substantially longer time of 3557~s, 
or 0.42~s per source. Therefore, the average running speed of the parametric fitting method is $>$10 times faster than that of the {\tt Mgformer}-based method. In practical applications, a trade-off between performance and time must be achieved regarding the 
% we recommend selecting the method based on the 
available computational resources and the amount of processed data each day.

\begin{deluxetable}{ccccc}[htbp]
\tablewidth{0.9\textwidth}
\tablecaption{Parametric fitting vs. {\tt Mgformer}-based method}\label{table:fit_vs_ml}
\tablehead{
\colhead{} & \colhead{Parametric fitting} & \colhead{{\tt Mgformer}}
}
\startdata
Run independently & \ding{51} & \ding{51}\\
Assign quantified scores & \ding{55} & \ding{51}\\
Multi-class classification & \ding{55}  &  \ding{51} \\
%Distinguish in early stage & \ding{55} & \ding{51} \\
Speed per source per core & 0.024 s & 0.42 s\\
%Flexibility  & Low & High \\
Execution sequence & Former & Latter \\
\enddata
\end{deluxetable}

\subsection{Special TDE types for future updates}

Our model training heavily relies on current spectroscopically identified transients. This ensures that our filter can effectively capture TDE candidates that share similar light curves as the typical TDEs. From a contrary view, this also implies that atypical TDEs will not be selected by our current filter, such as TDEs occurring in AGNs (e.g., \citealt{ps16dtm,AGN-TDE2}) and optical-bright repeating partial TDEs (e.g., \citealt{Payne2021,Somalwar_2025,at2022dbl}). In future updates, we will explore involving these TDE types or building specialized filters for them.

\section{Conclusion} \label{sec:conclusion}

Motivated by the prospect and scientific goal of finding TDEs with WFST, we have developed an automated classifier ({\tt TTC}) that can select 
% an integrated scheme for screening 
TDE candidates from multi-band light curve data with high recall and precision. It is a combination of two modules: Light curve parametric fitting and a Transformer({\tt Mgformer})-based classification network.
% , providing an effective and automated tool for identifying TDE candidates. 
These two modules can also work independently. We evaluate their individual performance on a ZTF light curve dataset of 7413 transients found between 2019 and 2024. The metrics are listed below:
\begin{tight_enumerate}    
    \item The parametric fitting method achieves a recall of 0.72 and a precision of 0.40.
    \item The \texttt{Mgformer}-based method achieves a recall of 0.79 and a precision of 0.76.
\end{tight_enumerate}

The \texttt{Mgformer}-based method is superior in the performance, yet much slower than the parametric fitting method (Section \ref{sec:CE}). The setup of modules allows a flexible trade-off between performance and time given the available computational resources and the amount of processed data each day. 

To enhance the ability of finding TDEs at early stages, data augmentation has been employed (Section \ref{sec:mlpreprocessing}). We have proved {\tt TTC}'s ability in finding TDE candidates within $\sim$30 days since the first discovery (Figure \ref{fig:time_vs_recall_precision}). Regarding the depth of WFST and ZTF ($\gtrsim$22 mag vs. $\gtrsim$20 mag), this enables WFST to capture a TDE candidate earlier than ZTF.

To test {\tt TTC}'s capability in finding TDE candidates in real time, we apply it to a set of 129 ZTF transients first detected in 2025 and have at least one detection in recent 10 days , and accurately find four TDEs among them, including ZTF25aajjeon (TDE 2025hbw) four days before it was classified as a featureless TDE at $z=0.6278$. This result demonstrates {\tt TTC}'s capability in finding distant TDEs.

Applying this method to the WFST survey dataset, we identify about 20 TDE candidates, including two highly confident candidates: WFST J134750.34+025201.4 and WFST J015306.85+065606.7. Unfortunately, neither of them have been bright enough for our spectroscopic follow-ups since they were discovered. Nonetheless, these two candidates demonstrate the capabilities of WFST in detecting potential TDEs in faint and distant galaxies through {\tt TTC}. We expect the situation would be greatly improved once the WFS differential database is ready, as it covers a $\sim$10 times larger sky area, leading to a $\sim$10 times higher anticipated discovery rate. 

In the future, following the commencement of the full-scale sky survey, a substantial number of TDE candidates with well-sampled photometry in the $u$ band are expected to be collected. The $u$-band plays a particularly important role in the photometric identification of TDEs due to its sensitivity to the blue continuum. Incorporating high-quality $u$-band data is anticipated to significantly enhance the model’s screening capability and improve the reliability of TDE classification.

The code developed for this study is open source and available at \url{https://github.com/Tico-Astro/TTC}.

\begin{acknowledgments}
We sincerely thank the anonymous referee for providing useful comments that significantly improve the quality of this article. This work is supported by the National Science Foundation of China (NSFC, Grant No. 12233008), the National Key R\&D Program of China (2023YFA1608100), the Strategic Priority Research Program of the Chinese Academy of Sciences (Grant No. XDB0550200), the Cyrus Chun Ying Tang Foundations, and the 111 Project for "Observational and Theoretical Research on Dark Matter and Dark Energy" (B23042).
\end{acknowledgments}

%WFST:2.5m

\vspace{5mm}
\facilities{WFST(2.5m); P48(1.2m)}

\software{\texttt{Astropy}\citep{astropy},  
          \texttt{Scipy}\citep{scipy},
          \texttt{Lasair} \citep{10.1093/rasti/rzae024-Lasair},
          \texttt{Pytorch}\citep{pytorch1,pytorch2},
          \texttt{Matplotlib}\citep{matplotlib},
          \texttt{Mgformer}\citep{MgFormer},
          \texttt{GpyTorch}\citep{Gpytorch}
          }

\bibliography{sample631}{}
\bibliographystyle{aasjournal}

\clearpage

\appendix

%\citet{Hammerstein_2023_ZTF_I}
%\citet{tde-catalog-2}
%this work

\section{TDE Samples Used for Training}

\startlongtable 
\begin{deluxetable*}{ccc}
\tablenum{A1}
\tablecaption{50 TDEs and Their ML Scores\label{tab:tdes_ml_scores}}
\tablewidth{0pt}
\tablehead{
\colhead{\textbf{\citet{Hammerstein_2023_ZTF_I}}} & \colhead{\textbf{\citet{tde-catalog-2}}} & \colhead{\textbf{Samples from TNS}}
}
\startdata
\texttt{ZTF18abxftqm $|$ 0.9736} & \texttt{ZTF19aakswrb $|$ 0.9323}   & \texttt{ZTF22aaabovl $|$ 0.9482} \\
\texttt{ZTF18acaqdaa $|$ 0.9782} & \texttt{ZTF19aaniqrr $|$ 0.9833} & \texttt{ZTF22aaaedas $|$ 0.9503} \\
\texttt{ZTF18acnbpmd $|$ 0.9941} & \texttt{ZTF20abgwfek $|$ 0.4769} & \texttt{ZTF22aaahtqz $|$ 0.9201} \\
\texttt{ZTF19aabbnzo $|$ 0.9903} & \texttt{ZTF20achpcvt $|$ 0.9741} & \texttt{ZTF22aacgcwv $|$ 0.0700} \\
\texttt{ZTF19aakiwze $|$ 0.9433} & \texttt{ZTF20acnznms $|$ 0.8265} & \texttt{ZTF22aadesap $|$ 0.0787} \\
\texttt{ZTF19aapreis $|$ 0.9893} & \texttt{ZTF20acwytxn $|$ 0.9512} & \texttt{ZTF22aagvrlq $|$ 0.8859} \\
\texttt{ZTF19aarioci $|$ 0.7062} & \texttt{ZTF21aaaokyp $|$ 0.6749} & \texttt{ZTF22aavvqyh $|$ 0.3056} \\
\texttt{ZTF19abhhjcc $|$ 0.9344} & \texttt{ZTF21aakfqwq $|$ 0.8968} & \texttt{ZTF22abajudi $|$ 0.9419} \\
\texttt{ZTF19abidbya $|$ 0.1680} & \texttt{ZTF21aanxhjv $|$ 0.8643} & \texttt{ZTF22abegjtx $|$ 0.3682} \\
\texttt{ZTF19abzrhgq $|$ 0.0101} & \texttt{ZTF21abaxaqq $|$ 0.6000} & \texttt{ZTF22abkfhua $|$ 0.9638} \\
\texttt{ZTF19accmaxo $|$ 0.9855} & \texttt{ZTF21abcgnqn $|$ 0.7749} & \texttt{ZTF23aadcbay $|$ 0.4717} \\
\texttt{ZTF19acspeuw $|$ 0.8991} & \texttt{ZTF21abhrchb $|$ 0.0720} & \texttt{ZTF23abcvbqq $|$ 0.9733} \\
\texttt{ZTF20aabqihu $|$ 0.8410} & \texttt{ZTF21abjrysr $|$ 0.6255} & \texttt{ZTF24abmybnp $|$ 0.6760} \\
\texttt{ZTF20aamqmfk $|$ 0.9944} & \texttt{ZTF21abqhkjd $|$ 0.9129} & \nodata \\
\texttt{ZTF20abefeab $|$ 0.9875} & \texttt{ZTF21abxngcz $|$ 0.9949} & \nodata \\
\texttt{ZTF20abfcszi $|$ 0.8230} & \texttt{ZTF21acafvhf $|$ 0.9777} & \nodata \\
\texttt{ZTF20abjwvae $|$ 0.9925} & \nodata & \nodata \\
\texttt{ZTF20abnorit $|$ 0.9829} & \nodata & \nodata \\
\texttt{ZTF20abowque $|$ 0.9901} & \nodata & \nodata \\
\texttt{ZTF20acitpfz $|$ 0.0206} & \nodata & \nodata \\
\texttt{ZTF20acqoiyt $|$ 0.9453} & \nodata & \nodata \\
\enddata
\tablecomments{We present the TDE sources used to train the ML model. Of these, 21 sources are from the ZTF-I TDE catalog \citep{Hammerstein_2023_ZTF_I}, 16 sources are from \citet{tde-catalog-2}, and the remaining 13 sources were selected based on spectral classifications from the TNS. The ML scores of them are provided.}
\label{ZTF-TDE-score}
\end{deluxetable*}

\section{Visualization on the Selection Cuts of Parametric Fitting}
\begin{figure*}[htbp]
\centering
\figurenum{B1}
\includegraphics[width=1.0\textwidth]{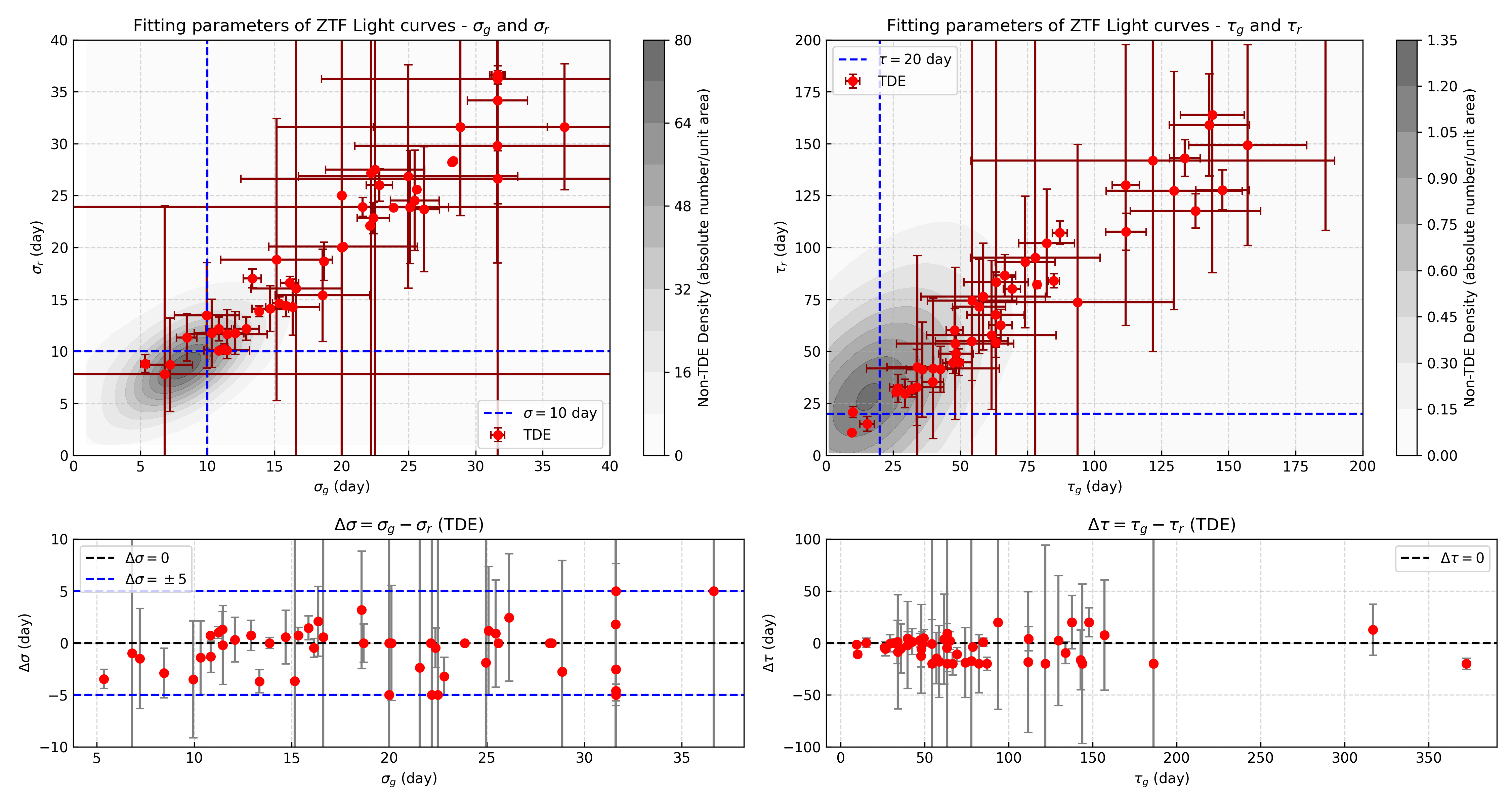}
\caption{Distribution of the fitted parameters. The parameters $\sigma$ and $\tau$ represent the characteristic rise and decay timescales, respectively, as derived from the adopted fitting function. The gray shaded contour regions represent the distribution of non-TDE sources, while TDEs are indicated by red data points. The upper panels show the correlations between $\sigma$ and $\tau$ for TDE and non-TDE samples, whereas the lower panels display the distributions of $\Delta\sigma = \sigma_g - \sigma_r$ and $\Delta\tau = \tau_g - \tau_r$ for TDEs. The blue dashed lines in the upper panels denote the reference thresholds for $\sigma$ and $\tau$, and the lower panels mark the limited fitting threshold for $\Delta\sigma$.
\label{fig:parameters_distribution}}
\end{figure*}

In \autoref{fig:parameters_distribution}, we plot the distributions of the fitted parameters for all samples. The shaded contour regions illustrate the parameter distribution of non-TDE sources, while red markers denote the TDE population. It is evident that TDEs typically exhibit rising timescales $\sigma$ exceeding approximately 10 days and declining timescales $\tau$ generally greater than 20 days.

\begin{figure*}[htbp]
\centering
\figurenum{B2}
\includegraphics[width=0.7\textwidth]{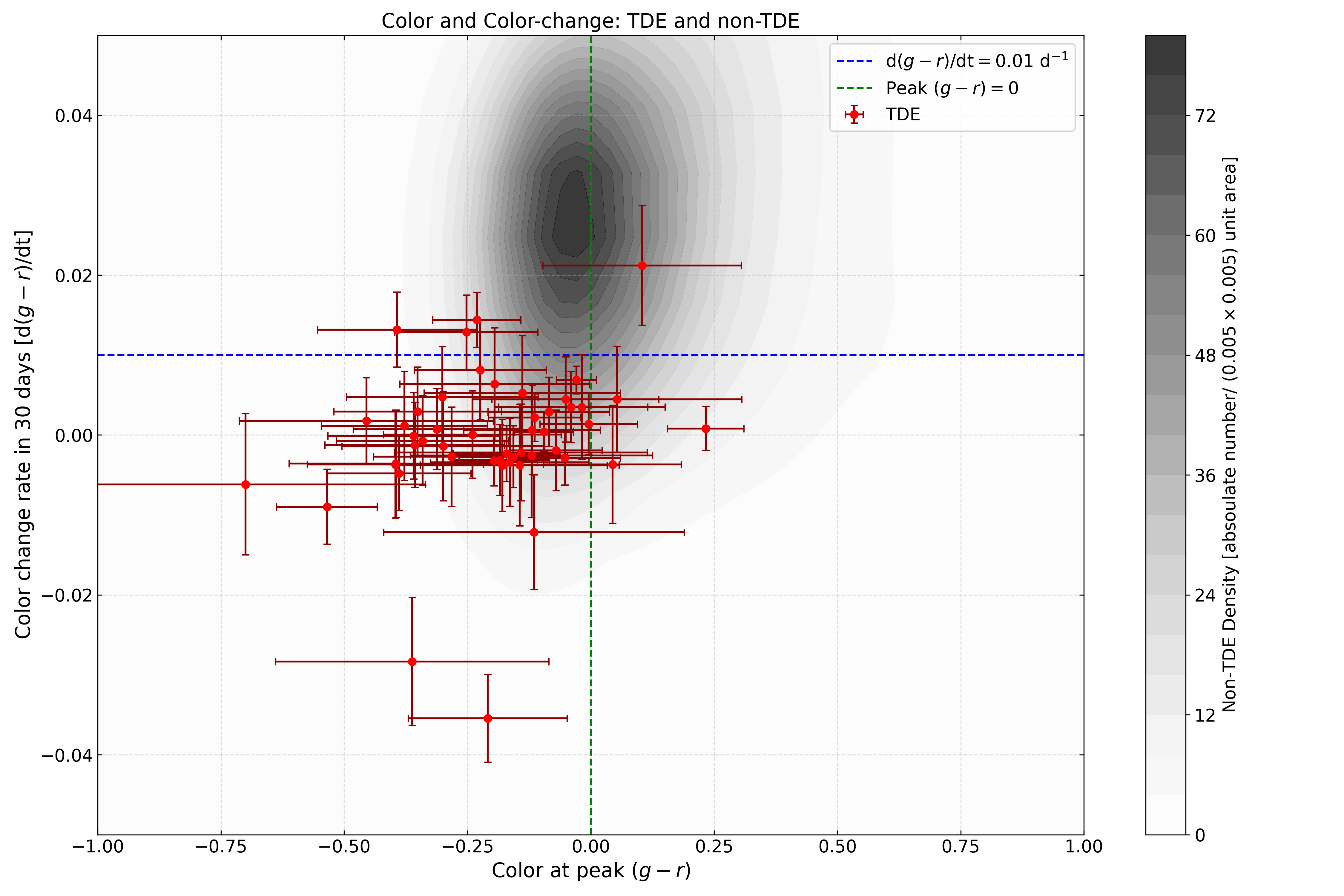}
\caption{Peak colors $(g-r)$ and color changes within 30 days after the peak for TDEs and non-TDEs. The green dashed line marks $g-r = 0$, and the blue dashed line indicates $\mathrm{d}(g-r)/\mathrm{dt}$. The gray background shows the distribution of non-TDEs, while the red points represent TDEs. The two dashed lines divide the plot into four quadrants, with most TDEs located in the third quadrant.
\label{fig:color_parameters_distribution}}
\end{figure*}

\autoref{fig:color_parameters_distribution} presents the peak colors of TDEs and non-TDEs, together with the color evolution within 30 days after the peak. It is evident that the vast majority of TDEs exhibit bluer peak colors and relatively slower color changes following maximum light. 
\vspace{10pt}

\section{Bonus Ability - Classify Supernovae}\label{sec:classifySN}

% As our primary objective is the identification of TDE candidates, 
Supernova classification is not the central focus of this study. Nonetheless, we observe that the model performs well when applied to supernova classification in WFST data.

We obtained 137 suspected SN candidates through the function fitting criterion without color and time scale restrictions which has peak difference flux $>0.05 \mathrm{\ mJy}$ in both $g$-band and $r$-band, and then we performed a cross-match with 2$^{\prime\prime}$ between the coordinates of supernovae reported in the TNS and the WFST database as of May 10, 2025. This resulted in the identification of 35 supernovae with well-sampled light curves in WFST—defined as having at least five photometric measurements in both the $g-$ and $r-$bands, and confirmed by spectroscopic classification. Among these, 25 are SN Ia, 9 SN II, and 1 SN Ic.

The verification results showed that the correct selection rate for SN Ia is 23/25, while the correct selection rate for SN II is 9/9, for SN Ib/c is 1/1. These results are consistent with our expectations. Notably, none of the sources were misclassified as TDE, and all exhibited TDE scores below 0.03. Therefore, we conclude that although the model is trained on photometric data from ZTF, its applicability can also extend to WFST data.




\end{document}